\documentclass[12pt,a4paper]{article}
\usepackage{amscd,amsmath,amsfonts,amssymb}
\usepackage[dvips]{epsfig}

\newcommand{\proj}{\mc{P}_{\!\!\omega}}
\newcommand{\embin}{{\hookrightarrow}}
\newcommand{\GL}{{\text{GL}}}

\newcommand{\Complex}{\mathbb{C}}
\newcommand{\Integer}{\mathbb{Z}}
\newcommand{\Natural}{\mathbb{N}}

\newcommand{\smat}{\begin{smallmatrix}}
\newcommand{\stam}{\end{smallmatrix}}
\newcommand{\mat}{\begin{pmatrix}}
\newcommand{\tam}{\end{pmatrix}}
\newcommand{\tr}{\text{tr}}

\newcommand{\mf}{\mathfrak} % Text in frak-style
\newcommand{\mc}{\mathcal} % Text in cal-style
\def\Inv{\text{\rm Inv}} 
\def\Iso{\Psi}
\def\iG{G^\omega} % The invariant subgroup
\def\sg{\mf{h}^\omega} % The relevant subalgebra
\def\ig{\mf{g}^\omega} % The invariant subalgebra
\def\bL{\mc{B}} % The symbol for the set of boundary labels
\def\areps{P_k^+} % Set of affine reps
\def\reps{P^+} % set of reps
\def\tareps{\bL^\omega_k} % Set of affine fractional sym. weights
\def\treps{\bL^\omega} % Set of fractional sym weights
\def\Hom{{\text{\rm Hom}}} % Symbol for set of homomorphisms
\def\End{{\text{\rm End}}} % symbol for set of endomorphisms
\def\Mat{{\text{\rm Mat}}} % symbol for set of matrices
\def\tT{{\rm T}} % Lie algebra generator
\def\tR{{\rm R}} % Representation matrix
\def\tK{{\rm K}} % Doubly covariant function

\def\U{{\mathcal U}}
\def\tCS{{\rm CS}}
\def\cS{{\mathcal S}}
\def\tF{{\rm F}}
\def\tf{{f}}
\def\tA{{\rm A}} 
\def\tL{{\rm L}}
\def\astk{\, , \, }

\def\om{\omega}
\def\QC{\mathbb{C}}
\def\asg{{\hat{\mf{g}}}}
\def\bz{{\bar z}}
\def\pl{\partial}
 
\def\bJ{{\bar J}} 
\def\bpl{\bar \partial} 
\def\bT{{\bar T}}
\def\Ad{{\rm Ad}}
\def\ad{{\rm ad}}
\def\id{{\rm id}}
\def\st{{\mathcal T}}
\def\C{{\mathcal C}} 
\def\rank{{\rm rank\,}}
\def\QZ{\mathbb{Z}}
\def\bZ{\QZ}
\def\a{\alpha}
\def\b{\beta}

\title{\bf Non-commutative gauge theory \\[3mm]  
 \bf of twisted D-branes}

\author{{\sc Anton Yu.\ Alekseev}\\[3mm] 
Universit{\'e} de Gen{\`e}ve, Section de math{\'e}matiques\\ 
2-4, rue du Li{\`e}vre, CH 1211 Gen{\`e}ve 24, Switzerland\\
\& ITP Uppsala University, S--75108 Uppsala, Sweden\\[2mm]
{\em Anton.Alekseev@math.unige.ch} \vspace*{3mm}
\and
{\sc Stefan Fredenhagen, Thomas Quella}\\
 {\sc and Volker Schomerus} \\[3mm] 
MPI f{\"u}r Gravitationsphysik, Albert-Einstein-Institut\\
Am M{\"u}hlenberg 1, D-14476 Golm, Germany\\[2mm]  
{\em stefan@aei.mpg.de}\quad 
{\em quella@aei.mpg.de} \\
 {\em vschomer@aei.mpg.de}}

\date{\medskip\small September 19, version to appear in Nucl.\ Phys.\ B}
  
\begin{document}

\maketitle
\baselineskip16pt 
  
\begin{abstract}
In this work we propose new non-commutative gauge theories that 
describe the dynamics of branes localized along twisted conjugacy 
classes on group manifolds. Our proposal is based on a careful 
analysis of the exact microscopic solution and it generalizes 
the matrix models (``fuzzy gauge theories'') that are used to 
study e.g.\  the bound state formation of point-like branes in a 
curved background. We also construct a large number of classical 
solutions and interpret them in terms of condensation processes 
on branes localized along twisted conjugacy classes.
\end{abstract}

\vspace*{-20cm}\noindent AEI-2002-027\hfill hep-th/0205123

\newpage
  
% -----------------------------------------------------------------------
% -----------------------------------------------------------------------
% ---------------------------------------------------n--------------------
\section{Introduction}

Branes on group manifolds have received a lot of attention 
throughout the last years. Even though the group manifold 
of $SU(2) \cong S^3$ is the only one that can appear directly 
as part of a string background (e.g.\ in the background of 
NS5-branes), other group manifolds provide simple toy models 
for studying the behavior of branes in curved backgrounds. In 
fact, the symmetry of group manifolds allows to control string 
perturbation theory beyond the supergravity regime and hence 
it renders investigations of brane spectra and brane dynamics 
at finite curvature possible. Moreover, group manifolds are the 
starting point for coset and orbifold constructions and thereby 
allow to obtain many less symmetric string backgrounds.%
\smallskip

Open strings on group manifolds are modeled with the help 
of the Wess-Zumino-Witten (WZW) theory on the upper half 
$\Sigma = \{ \Im z \geq 0\}$ of the complex plane. Boundary 
conditions giving rise to a consistent open string spectrum 
were first proposed by Cardy \cite{Cardy:1989ir} but their 
geometric interpretation in terms of brane geometry was only 
uncovered much later in \cite{Alekseev:1998mc}. There it was 
shown that Cardy's boundary theories describe branes that 
wrap certain integer conjugacy classes on the group manifold 
$G$ (see also \cite{Gawedzki:1999bq,Stanciu:1999id}). 
\smallskip

The Born-Infeld action has been used in \cite{Bachas:2000ik,
Pawelczyk:2000ah} to explain the stability of such branes in 
a weakly curved background (see also \cite{Bordalo:2001ec}
for more general cases). Once several branes are placed on 
a group manifold, they tend to form new bound states. These 
dynamics are described by non-commutative gauge theories on 
the quantized (``fuzzy'') conjugacy classes. The latter were 
derived from the exact boundary conformal field theory in 
\cite{Alekseev:1999bs}. Even though these actions are 
applicable only at large level $k$, many of the condensation 
processes they encode are known to possess a deformation to 
finite $k$ (see \cite{Alekseev:2000jx,Fredenhagen:2000ei}).     
\medskip
  
Branes localized along conjugacy classes are not the 
only ones that admit an exact solution. In \cite{Birke:1999ik} 
Birke, Fuchs and Schweigert constructed the open string 
spectrum for new sets of branes that were shown later 
to be localized along so-called twisted conjugacy 
classes \cite{Felder:1999ka}. These extra series of branes 
come associated with non-trivial outer automorphisms 
$\omega_G$ of the group $G$. 
\medskip

In the present work we aim at understanding the dynamical 
properties of such twisted D-branes on group manifolds.  
After a brief introduction to the geometry and conformal 
field theory of maximally symmetric branes on group 
manifolds, we will use the microscopic information on 
open string spectra to construct the space of functions 
on all these branes in the limit where the level $k$ is 
sent to infinity. We show that the space of functions 
can be equipped with an associative and non-commutative 
product. For non-trivial twists $\omega_G \neq \id$, the 
geometry of these branes is not just a simple matrix 
geometry as in the case of branes on conjugacy classes
but is based on a certain algebra of {\em matrix valued
functions} on the group.
The associated gauge theories are then easily obtained by 
copying the computations from \cite{Alekseev:1999bs}. In 
the final section we analyze classical solutions. Most 
importantly, we shall establish that for a given 
twist $\omega_G$, all the branes appear as bound 
states from stacks of one distinguished elementary 
twisted brane. This is analogous to the situation of branes 
wrapping ordinary conjugacy classes which can all be 
built from condensates of point-like branes 
\cite{Alekseev:2000fd} (see also \cite{Hashimoto:2001xy,
Hikida:2001py}) by some analogue of Myers dielectric 
effect \cite{Myers:1999ps}. Let us remark 
that the new condensation processes we find are  
consistent with the charge conservation laws  
proposed in \cite{Fredenhagen:2000ei}.

% -----------------------------------------------------------------------
% -----------------------------------------------------------------------
% -----------------------------------------------------------------------
\section{D-branes on group manifolds}

This section is devoted to the description of maximally 
symmetric branes on group manifolds. Following \cite{Alekseev:1998mc,
Felder:1999ka}, we will begin with a brief review of their classical 
geometry. Then, in the second subsection, we shall present 
some basic results on the boundary conformal field theory 
of such branes.    

% -----------------------------------------------------------------------
% -----------------------------------------------------------------------
\subsection{The geometry of branes on group manifolds}

Strings on the group manifold of a simple and simply connected 
group~$G$ are described by the WZW-model. Its action is evaluated on 
fields $g: \Sigma \to G$ taking values in~$G$ and it involves 
one (integer) coupling constant $k$, which is known as the ``level''. 
For our purposes it is most convenient to think of $k$ as controlling 
the size (in string units) of the background. Large values of $k$ 
correspond to a large volume of the group manifold. When dealing 
with open strings at tree level, the 2-dimensional world sheet 
$\Sigma$ is taken to be the upper half plane $\Sigma = \{ z \in 
\QC | \Im z \geq 0\}$. 
\smallskip

Along the boundary of this world sheet we need to impose some 
boundary condition. Here we shall analyze boundary conditions 
that preserve the full bulk symmetry of the model, i.e.\ the 
affine algebra $\asg_k$. These boundary conditions are formulated 
in terms of the chiral currents 
\begin{equation}
 J (z) \ = \ k \, g^{-1}(z,\bz) \pl g(z,\bz ) \ \ \  , \ \ \ 
   \bJ (\bz) \ = \ -k \, \bpl g(z,\bz ) \, g^{-1}(z,\bz)\ \ .
\end{equation}
Note that $J$ and $\bJ$ take values in the finite dimensional 
Lie algebra $\mf{g}$ of the group~$G$. Along the real line we glue 
the holomorphic and the anti-holomorphic currents  according to 
\begin{equation}
  \label{glue}  
  J(z) \ = \ \Lambda \bJ(\bz)  \ \ \ 
  \text{ for all } \ \ \ z = \bz 
\end{equation}
where $\Lambda$ is an appropriate automorphism of the current  
algebra $\asg_k$ (see e.g.\ \cite{Recknagel:1998sb}). The choice of 
$\Lambda$ is restricted by the requirement of conformal 
invariance which means that $T(z) = \bT(\bz)$ all along the 
boundary. Here $T, \bT$ are the non-vanishing components
of the stress energy tensor. They can be obtained, as usual,
through the Sugawara construction.
\smallskip

The allowed automorphisms $\Lambda$ of the affine Lie algebra~$\asg_k$
are easily classified. They are all of the form 
\begin{equation}
  \label{Lam}
  \Lambda \ = \ \Omega \circ \Ad_g \ \ 
  \text{ for some } \  g \in G \ \ . 
\end{equation}
Here, $\Ad_g$ denotes the adjoint action of the group element 
$g$ on the current algebra $\asg_k$. It is induced in the 
obvious way from the adjoint action of~$G$ on the finite 
dimensional Lie algebra $\mf{g}$. The automorphism $\Omega$ 
does not come from conjugation with some element $g$. More 
precisely, it is an outer automorphism of the current 
algebra. Such outer automorphisms $\Omega = \Omega_\omega$ 
come with symmetries $\omega$ of the Dynkin diagram of the 
finite dimensional Lie algebra $\mf{g}$. One may show that 
the choice of $\omega$ and $g \in G$ in eq.~\eqref{Lam} 
exhausts all possibilities for the gluing automorphism
$\Lambda$ (see e.g.~\cite{Kac:1990}). 
\medskip

So far, our discussion of the admissible types of gluing 
automorphisms $\Lambda$ has been fairly abstract. It is 
possible, however, to associate some concrete geometry 
with each choice of $\Lambda$. This was initiated in 
\cite{Alekseev:1998mc} for $\omega = \id$ and extended to 
non-trivial symmetries $\omega \neq \id$ in \cite{Felder:1999ka} 
(see also \cite{Gawedzki:1999bq}, \cite{Stanciu:1999id}). 
\smallskip

Let us assume first that the element $g$ in eq.~\eqref{Lam}
coincides with the group unit $g = e$. This means that $\Lambda
= \Omega = \Omega_\omega$ is determined by $\omega$ alone. The
diagram symmetry $\omega$ induces an (outer) automorphism 
$\omega$ of the finite dimensional Lie algebra $\mf{g}$ through 
the unique correspondence between vertices of the Dynkin diagram 
and simple roots. After exponentiation, $\omega$ furnishes
an automorphism $\omega_G$ of the group~$G$. One can show that 
the gluing conditions~\eqref{glue} force the string ends to 
stay on one of the following $\omega$-twisted conjugacy 
classes
$$ C^\omega_u \ := \ \{\,  h u\,  \omega_G(h^{-1})\ |\ h \in G 
\, \} \ \ . $$
The subsets $C^\omega_u \subset G$ are parametrized 
by equivalence classes of group elements $u$ where the 
equivalence relation between two elements $u,v \in G$ is
given by: $u \sim_\omega v$ iff $v \in C^\omega_u$. Note 
that this parameter space $\U^\omega$ of equivalence classes
is not a manifold, i.e.\ it contains singular points.  
\smallskip

To describe the topology of $C^\omega_u$ and the parameter 
space $\U^\omega$ (at least locally), we need some more 
notation. By construction, the action of $\omega$ on 
$\mf{g}$ can be restricted to an action on the Cartan subalgebra $\st$. 
We shall denote the subspace of elements which are invariant 
under the action of $\omega$ by $\st^\omega \subset \st$. 
Elements in $\st^\omega$ generate a torus $T^\omega \subset G$. 
One may show that the generic $\omega$-twisted conjugacy class
$\C^\omega_u$ looks like the quotient $G/T^\omega$. Hence, 
the dimension of the generic submanifolds $\C^\omega_u$ 
is $\dim G - \dim T^\omega$ and the parameter space has
dimension $\dim \st^\omega$ almost everywhere.
In other words, there are $\dim \st^\omega$ 
directions transverse to a generic twisted conjugacy 
class. This implies that the branes associated with the 
trivial diagram automorphism $\omega = \id$ have the largest 
number of transverse directions. It is given by the rank of 
the Lie algebra. 
\smallskip 

As we shall see below, not all these submanifolds $C^\omega_u$ 
can be wrapped by branes on group manifolds. There exists some 
integrality requirement that can be understood in various ways, 
e.g.\ as quantization condition within a semiclassical analysis 
\cite{Alekseev:1998mc} of the brane's stability 
\cite{Bachas:2000ik, Pawelczyk:2000ah, Bordalo:2001ec}.
This implies that there is only a finite set of 
allowed branes (if $k$ is finite). The number of branes depends 
on the volume of the group measured in string units. 
\smallskip

Let us become somewhat more explicit for $G = SU(N)$. The 
simplest case is certainly $N=2$ because there exists no 
non-trivial diagram automorphism $\omega$. The conjugacy 
classes $C^\id_u$ are 2-spheres $S^2 \subset S^3 \cong 
SU(2)$ for generic points $u$ and they consist of a 
single point when $u = \pm e$ in the center of $SU(2)$.  
More generally, the formulas $\dim SU(N) = (N-1)(N+1)$ and 
$\rank SU(N) = (N-1)$ show that the generic submanifolds 
$C^\id_u$ have dimension $\dim C^\id_u = (N-1)N$.
In addition, there are singular cases, $N$ of which are associated
with elements $u$ in the center $\bZ_N \subset SU(N)$. The 
corresponding submanifolds $C^\id_u$ are 0-dimensional. 
Note that all the submanifolds $C^\id_u$ are even 
dimensional. Similarly, the generic manifolds $C^\omega_u$ 
for the non-trivial diagram symmetry $\omega$ have 
dimension $\dim C^\omega_u = (N-1)(N+1/2)$ for odd N 
and $\dim C^\omega_u = N^{2}-N/2-1$ whenever 
$N$ is even. For some exceptional values of 
$u$, the dimension can be lower.
For a complete discussion we refer the reader
to~\cite{Stanciu:2001vw}.
\medskip

So far we restricted ourselves to $\Lambda=\Omega_\omega$ being 
a diagram automorphism. As we stated before, the general case 
is obtained by admitting an additional inner automorphism of 
the form $\Ad_g$. Geometrically, the latter corresponds to rigid 
translations on the group induced from the left action of $g$ 
on the group manifold (see e.g.\ \cite{Recknagel:1998ih, Recknagel:1998ut}).
The freedom of 
translating branes on~$G$ does not lead to any new charges or
to essentially new physics and 
we shall not consider it any further, i.e.\ we shall assume 
$g = e$ in what follows. 

% -----------------------------------------------------------------------
% -----------------------------------------------------------------------
\subsection{\label{sc:CFT}The conformal field theory description}

  The branes we considered in the previous subsection may be described
through an exactly solvable conformal field theory. In particular, 
there exists a detailed knowledge about their open string spectra
based on the work of Cardy \cite{Cardy:1989ir} and of Birke, Fuchs 
and Schweigert \cite{Birke:1999ik}. 
\medskip

We shall use $\alpha,\beta,\ldots\in\tareps$ to label
different conformal boundary conditions of the boundary conformal 
field theories associated with the gluing conditions eq.~\eqref{glue}
on the currents. The set $\tareps$ depends on the choice of the diagram
automorphism~$\omega$ and on the level~$k$. For the trivial diagram
automorphism $\omega=\id$, $\bL^\id_k = \areps$ coincides with
the set of primaries of the affine Kac-Moody algebra $\asg_k$. 
The latter is well known to form a subset in the space~$\reps$ 
of dominant integral weights which label equivalence classes of 
irreducible representations for the finite dimensional 
Lie algebra $\mf{g}$. To keep notations simple, we will not 
distinguish in notation between elements of~$\reps$ and~$\areps$
and denote them both by capital letters $A,B,C,\ldots$

The automorphism $\omega$ generates a map $\omega_k:\areps
\rightarrow\areps$. In fact, given an irreducible representation 
$\tau$ of $\mf{g}$, we can define another representation by 
composition $\tau\circ\omega$. The class of $\tau \circ \omega$ 
is independent of the choice of $\tau \in [\tau]$ and so we obtain 
a map $\omega:\reps\rightarrow\reps$. The latter descends to 
the subset $\areps\subset\reps$. A weight $A\in\areps$ is said 
to be ($\omega$-)symmetric if it is invariant under the action of 
$\omega$, i.e.\ if $\omega(A)=A$. According to the results of 
\cite{Cardy:1989ir, Birke:1999ik}, the labels~$\alpha$ for 
branes associated with the diagram automorphism $\omega$ take 
values in a certain subset~$\tareps\subset\treps=\proj(P^+)$
of dominant fractional symmetric weights. Here $\proj =
\frac{1}{N}\bigl(1+\cdots+\omega^{N-1}\bigr) $ is the projection of weights 
onto their ($\omega$-)symmetric part with~$N$ denoting the order 
of~$\omega$. Finally let us briefly mention that the number of boundary
conditions~$\alpha\in\tareps$ is equal to the number of symmetric
weights in~$\areps$. On the other hand, $\tareps\neq\proj(P_k^+)$ 
in contrast to what one might expect naively. We will not need these 
details here and refer the interested reader to~\cite{Birke:1999ik} (see 
also~\cite{Quella:2002wi}). Our considerations below will mostly 
take place in the limiting regime of large level $k$ where we can 
identify $\treps_{\infty}=\treps=\proj(P^+)$. 
\smallskip

Before we continue to outline the conformal field theory 
results let us briefly summarize some conventions we will
be using in the limit~$k\rightarrow\infty$. In this case, 
the fractional symmetric weights which label boundary 
conditions can be described explicitly by
\begin{equation}
\treps\ =\ \Bigl\{\alpha=\sum\lambda_i\omega_i\:\Bigl|\:
\lambda_i=\lambda_{\omega(i)}\:,\: l_i\lambda_i\in\Natural_0\Bigr\}\ \ .
\end{equation}
The numbers~$l_i$ denote the length of the orbit of the
fundamental weights~$\omega_i$ under the automorphism~$\omega$.
A distinguished element of this set of fractional symmetric
weights is $\rho_\omega=(1/l_1,1/l_2,\cdots)\in\treps$. It can 
be considered as a twisted counterpart of the Weyl vector 
$\rho=(1,1,\cdots)\in\reps$. Under the assumption of infinite 
level~$k$, the representations of~$\mf{g}$ and~$\asg_k$ are 
both labeled by the same set~$\reps$. With the identifications 
of this section in mind, we will always assume that we fix the 
weights before we let the level run to infinity.
\medskip

Our main goal now is to explain the open string spectra that come
with the maximally symmetric branes on group manifolds. For a pair 
of boundary labels $\alpha,\beta\in\tareps$ associated with the 
same diagram automorphism $\omega$, the partition function is of 
the form
\begin{equation}
  \label{eq:CFTSpectrum}
  Z^\omega_{\a \b}(q) \ 
  = \ \sum_{A\in\areps} {\bigl(n_A^\omega\bigr)_\alpha}^\beta\,\chi_A (q) \ \ . 
\end{equation}
Here, $\chi_A(q)$ denote the characters of the current algebra 
$\asg_k$ and $\a,\b\in\tareps$. Their appearance in the expansion 
\eqref{eq:CFTSpectrum} reflects the fact that all the (twisted) 
conjugacy classes admit an obvious action of the Lie group~$G$ 
by (twisted) conjugation. Consistency requires the numbers 
${\bigl(n_A^\omega\bigr)_\alpha}^\beta$ to be non-negative integers. 
\smallskip

There exists a very simple argument due to Behrend et
al.~\cite{Behrend:1999bn, Behrend:1998fd} 
which shows that the matrices~${\bigl(n_A^\omega\bigr)_\alpha}^\beta$ give 
rise to a representation of the fusion algebra of~$\asg_k$. This 
means that they obey the relations 
\begin{equation}
 \label{nfus}
  \sum_{\b\in\tareps} {\bigl(n_A^\omega\bigr)_\alpha}^\beta \ 
    {\bigl(n_B^\omega\bigr)_\beta}^\gamma \ = \ 
    \sum_{C\in\areps} {N_{AB}}^{C} {\bigl(n_C^\omega\bigr)_\alpha}^\gamma\ \ , 
\end{equation}
where ${N_{AB}}^C$ are the fusion rules of the current algebra~$\asg_k$. 
The argument of~\cite{Behrend:1999bn, Behrend:1998fd} starts from a general
Ansatz for the boundary state assigned to $\alpha\in\tareps$.
Using world sheet duality, one can express the matrices~$n_A^\omega$ in 
terms of the coefficients of the boundary states and the modular 
matrix~$S$ for the current algebra~$\asg_k$. The general form 
of this expression is then sufficient to check the relations \eqref{nfus}
(see \cite{Behrend:1999bn, Behrend:1998fd} for details). 
\smallskip
 
The expression for the matrices $n_A^\omega$ that is given 
in~\cite{Birke:1999ik, Fuchs:1999zi} resembles the familiar 
Verlinde formula,  
\begin{equation}
  \label{annuluscf}
  {\bigl(n_A^\omega\bigr)_\alpha}^\beta \ = \ 
\sum_{\substack{\lambda\in\areps\\\omega(\lambda)=\lambda}}
\frac{\bar{S}^{\omega}_{\lambda\beta}S^{\omega}_{\lambda \alpha}
S_{\lambda A }}{S_{\lambda\, 0}} \ \ \ \text{ for } \ \ \  
\a,\b \in \bL^\om_k \ \ \text{ and } \ \ A \in\areps\ \ .  
\end{equation}
It contains a unitary matrix $S^{\omega}$ whose entries  
$S^\om_{\lambda \a}$ are indexed by two $\omega$-symmetric labels 
$\lambda\in\areps, \alpha\in\tareps$, i.e.\ they obey $\omega(\lambda)= 
\lambda$ and $\omega(\a)=\a$. When $\om = \id$, the matrix $S^\om$
coincides with the usual $S$-matrix so that Verlinde's 
formula~\cite{Verlinde:1988sn} implies  
$$ {\bigl(n_A^\omega\bigr)_\alpha}^\beta \ = \ 
{N_{A \a}}^{\b} \ \ \ \text{ for all }
   \ \ \a,\b,A\in\areps=\bL_k^\id\ \ .$$
This reproduces Cardy's results on the boundary partition
functions \cite{Cardy:1989ir}. For non-trivial automorphism $\om$, 
the matrix $S^\om$ describes modular transformations of 
twined characters. An explicit formula for $S^\om$ can be
found in~\cite{Birke:1999ik}
\begin{equation}
  S_{\lambda\alpha}^\omega \ \sim\ \sum_{w\in W_\omega}\ 
  \epsilon_\omega(w) \exp\Bigl(-\frac{2\pi i}{k+g^\vee}
  \bigl(w(\lambda+\rho),\alpha+\rho_\omega\bigr)\Bigr)\ \ .
\end{equation}
Here, $W_\omega \subset W$ is the $\omega$-invariant part in the Weyl group
of~$\mf{g}$. As~$W_\omega$ can be considered as the Weyl group of another
Lie algebra~\cite{Fuchs:1996zr} it comes with a natural sign
function~$\epsilon_\om$.    
\medskip 

There exists a generalized state-field correspondence that 
associates to every highest weight state~$A\in\areps$ in 
the Hilbert space~\eqref{eq:CFTSpectrum} of the $(\alpha,\beta)$
boundary conformal field theory a boundary 
primary field~$\psi_A^{(\a\b)}$ living between boundaries
$\alpha,\beta\in\tareps$. The general structure of the 
boundary operator product expansion (OPE) is given by
\begin{equation}
  \label{eq:BOPE}
  \psi_A^{(\alpha\beta)}(x)\, \psi_B^{(\beta\gamma)}(y)
  \ \sim\ \sum_{C}\, (x-y)^{h_C-h_A-h_B}
    \, C_{AB}^{(\alpha\beta\gamma)\:C}\psi_C^{(\alpha\gamma)}(y)
  \quad\text{ for }\quad x<y
\end{equation}
where the numbers~$h_A$ denote the conformal weights of the
fields which, in the case at hand, are given by
\begin{equation}
  \label{eq:ConformalWeights}
  h_A\ =\ \frac{C_A}{2(k+g^\vee)}\ \ .
\end{equation}
Here, $C_A$ is the quadratic Casimir of the representation~$A\in\areps$
(see eq.~\eqref{eq:QuadraticCasimir} below) and~$g^\vee$ denotes the
dual Coxeter number of~$\mf{g}$. For a consistent conformal 
field theory, the 
structure constants~$C_{AB}^{(\alpha\beta\gamma)C}$ have to 
satisfy so-called sewing constraints~\cite{Lewellen:1992tb}
(see also \cite{Pradisi:1996yd,Runkel:1998pm,Runkel:1999dz}). 
One of these constraints expresses the associativity of the OPE, 
\begin{equation}
  \label{eq:SewingConstraints}
  C_{AB}^{(\alpha\beta\gamma)E}\, C_{CD}^{(\gamma\delta\alpha)E^+}
  \, C_{EE^+}^{(\alpha\gamma\alpha)0}
  \ =\ \sum_H\,  C_{BC}^{(\beta\gamma\delta)H}\, 
  C_{AH}^{(\alpha\beta\delta)D^+}\, C_{D^+D}^{(\alpha\delta\alpha)0}
  \quad\tF_{HE}\left[\begin{smallmatrix}B&C\\A&D\end{smallmatrix}
\right]\quad
\end{equation}
where the symbol ${\rm F}$ denotes the fusing matrix. For
the $\asg_k$-WZW model, this fusing-matrix is closely related 
to the $6j$--symbol of the corresponding quantum group at 
$(k+g^\vee)^{th}$ root of unity. In the limit~$k\to\infty$, it 
thus reduces to the $6j$--symbol of~$\mf{g}$. Solutions of 
the sewing constraints for the standard case $\omega_G = \id$ 
can be found in~\cite{Runkel:1998pm, Alekseev:1999bs, Felder:1999ka, 
Behrend:1999bn}. The boundary OPE for non-trivially twisted 
branes, on the other hand, is not yet known. Our results
below implicitly contain these solutions in the limit 
$k \rightarrow \infty$.  
\medskip

The spectrum of ordinary conjugacy classes can be explained 
in detail. For simplicity, we shall restrict to 
$G = SU(2)$. In this case, generic conjugacy classes are 
2-spheres and the space of functions thereon is spanned 
by spherical harmonics $Y^{j/2}_m, |m| \leq j/2$ and $j = 
0,2,4,\dots$\footnote{To be consistent with our 
treatment of other groups below, we use a convention in which 
the representations are labeled by Dynkin labels rather than spins.}  
The space of spherical harmonics is precisely 
reproduced by ground states in the boundary theory 
$\alpha$ when we send $\a$ (and hence $k$) to infinity. 
For finite $\a$, the angular momentum $j$ is cut off at a 
finite value $j = \min(2\a, 2k-2\a) \leq 2 \a$. This means
that the brane's world-volume is ``fuzzy'', since resolving 
small distances would require large angular momenta. The
relation between branes on $SU(2)$ and the familiar non-%
commutative fuzzy 2-spheres \cite{Hoppe:1989gk, Madore:1992bw}
was fully analyzed 
in~\cite{Alekseev:1999bs} and it provides a very important  
example of an open string non-commutative geometry that goes 
beyond the familiar case of branes in flat space~\cite{Douglas:1998fm,
Chu:1998qz, Schomerus:1999ug}. 
The analysis of \cite{Alekseev:1999bs} goes much beyond the study
of partition functions as it employs detailed information
on the operator product expansions of open string vertex
operators based on \cite{Runkel:1998pm}. Using the results in 
\cite{Felder:1999ka, Felder:1999cv, Felder:1999mq} it is
easy to generalize all these remarks 
on ordinary conjugacy classes to other groups (see also 
\cite{Hoppe:1989gk} for more details and explicit formulas on 
fuzzy conjugacy classes).  
\smallskip

Twisted conjugacy classes are more difficult to understand. 
This is related to the fact that they are never ``small''. More
precisely, it is not possible to fit a generic twisted conjugacy 
class into an arbitrarily small neighborhood of the group identity 
unless the twist $\omega$ is trivial. This implies that the 
spectrum of angular momenta in $Z_{\a\a}^\om$ is not cut off 
before it reaches the obvious large momentum cut-off that is 
set by the volume of the group, i.e.\ by the level $k$. For
large $\a\in\tareps$ (and large $k$) the ground states in the boundary 
theory span the space of functions on the generic twisted conjugacy 
classes $C^\om_u$ \cite{Felder:1999ka}. The non-commutative geometry 
associated with twisted conjugacy classes with finite $\a$, however, 
was unknown and it is the main subject of the next section.     

% -----------------------------------------------------------------------
% -----------------------------------------------------------------------
% -----------------------------------------------------------------------
\section{Non-commutative geometry and gauge theory}

The existing information on the open string spectra of twisted 
D-branes on group manifolds can be turned into a proposal for 
the algebra of functions on these branes. The algebras turn out 
to be non-commutative due to the presence of a non-vanishing B-field. 
In the second subsection we shall spell out gauge theories that 
encode the dynamics of twisted D-branes.     

% -----------------------------------------------------------------------
% -----------------------------------------------------------------------
\subsection{The world-volume algebra}

According to the procedure suggested in \cite{Schomerus:1999ug,
Alekseev:1999bs} (see also \cite{Frohlich:1993es,
Frohlich:1995mr} 
for similar proposals in case of closed strings), the world-volume 
geometry of branes can be read off from the correlators of boundary 
operators in the decoupling regime~$k \rightarrow \infty$
\footnote{When taking the limit $k\to \infty $, we keep the open string
data stable. In particular, the low energy spectrum of open string modes
does not change. A limit where the closed string data is kept stable
instead has been considered in~\cite{Felder:1999ka}.}. Note that 
the conformal dimensions~\eqref{eq:ConformalWeights} of the boundary 
fields vanish in this limit so that the operator product expansion
\eqref{eq:BOPE} becomes independent of the world sheet coordinates.
In particular, all conformal families in the boundary theory contribute 
to the massless sector and thus to the gauge theory which governs the 
low energy dynamics of the D-branes we are studying. The program we 
have just sketched has been carried out successfully for the 
untwisted branes on compact group manifolds. In case of~$A_1$, 
this leads to the well known fuzzy spheres~\cite{Madore:1992bw, 
Alekseev:1999bs}. We will now describe the generalization to 
arbitrary $\omega$-twisted D-branes on compact simply-connected
simple group manifolds~$G$. Note that all simple Lie groups of 
ADE~type (except from $A_1$) admit non-trivial outer automorphisms.

\begin{table}
\centerline{\begin{tabular}{cccccc}
  $\mf{g}$ & order & $\ig$ & $x_e$ & $G$ & $\iG$ \\\hline
  $A_2$ & 2 & $A_1$ & 4 & $SU(3)$ & $SO(3)$ \\
  $A_{2n-1}$ & 2 & $C_n$ & 1 & $SU(2n)$ & $Sp(2n,\Complex)\cap SU(2n)$ \\
  $A_{2n}$ & 2 & $B_n$ & 2 & $SU(2n+1)$ & $SO(2n+1)$ \\
  $D_4$ & 3 & $G_2$ & 1& $Spin (8)=\widetilde{SO}(8)$ & $\widetilde{G}_2$ \\
  $D_n$ & 2 & $B_{n-1}$ & 1& $Spin (2n)=\widetilde{SO}(2n)$ & $Spin (2n-1)
  =\widetilde{SO}(2n-1)$ \\
  $E_6$ & 2 & $F_4$ & 1 & $\widetilde{E}_6$ & $\widetilde{F}_4$
\end{tabular}}
  \caption{\label{tb:Overview}Simple Lie algebras, groups and data
           related to outer automorphisms.}
\end{table}

In section~\ref{sc:CFT} we explained that all possible 
$\omega$-twisted D-branes are labeled by the set~$\treps$ of 
fractional symmetric weights of~$\mf{g}$. In this section we 
propose an alternative in which the same boundary conditions 
are labeled by representations~$\reps_{\iG}$ of the invariant
subgroup~$\iG=\{g\in G\:|\:\omega(g)=g\}$. Let us stress that 
we use representations of the group $\iG$ and not of its Lie
algebra. The two sets of representations agree only if
$\iG$ is simply-connected. This is the case for all Lie algebras 
but the $A_{2n}$ series where $\iG=SO(2n+1)$ and $\reps_{\iG}=
\bigl\{A\in\reps_{\iG}\bigl|A_n\text{ even}\bigl\}$. An overview 
over the relevant groups and Lie algebras can be found 
in Table~\ref{tb:Overview}. It is far from obvious that such a
labeling of twisted boundary conditions through representations 
of the invariant subgroup exists. But as we explain in 
Appendix~\ref{sc:Correspondence} one can indeed construct a 
structure preserving map $\Iso:\reps_{\iG}
\rightarrow\treps$ between the set of irreducible representations 
of~$\iG$ and the boundary labels for $\omega$-twisted branes.
For this reason, we will henceforth use both kinds of labels
$a=\Iso^{-1}(\alpha),b=\Iso^{-1}(\beta),\ldots$ 
equivalently when referring to boundary conditions\footnote{The map
$\Iso $ specifies the way in which we treat the boundary labels 
while sending $k$ to infinity. We refer the reader to
Appendix~\ref{sc:Correspondence} for details.}.
\smallskip 

Once the existence of $\Iso$ is established, we can 
formulate our proposal for the world-volume algebra. To this 
end, let~$V_a,V_b$ be two representation spaces for irreducible
representations~$a,b\in\reps_{\iG}\cong\treps$ of $\iG$. As we
will argue below, the relevant algebraic structure governing 
strings stretching between two D-branes of type~$a$ and~$b$,
respectively, is given by 
\begin{equation}
  \label{eq:DBraneModule}
  \mc{A}^{(a,b)}\cong  \Inv_{\iG}\Bigl(\mc{F}^{(a,b)}\Bigr) \ \ \
  \text{ where } \ \ \ 
  \mc{F}^{(a,b)}\ :=\ \mc{F}(G)\otimes \Hom(V_a,V_b)\ \ .
\end{equation}
Here $\mc{F}(G)$ denotes the algebra of functions on the group~$G$
and $\Hom(V_a,V_b)$ is the vector space of linear transformations 
from~$V_a$ to~$V_b$. The auxiliary space $\mc{F}^{(a,b)}$ can be 
regarded as a vector space of matrix valued functions on the 
group~$G$. It carries an action of the product group $G\times\iG$
defined by 
\begin{equation}
  \tA^{(g,h)}(g^\prime) \ =\ \tR_b(h)\tA(g^{-1}g^\prime h)
  \tR_a(h)^{-1},
\end{equation}
where $\tR_a(h)\in\GL(V_a),\tR_b(h)\in\GL(V_b)$ are 
the representation matrices of~$h$. In our construction of 
$\mc{A}^{(a,b)}$, we restrict to matrix valued functions
$\Inv_{\iG}\bigl(\mc{F}^{(a,b)}\bigr)$
which are invariant under the action of $\{\id\}\times\iG\subset G \times \iG$.
Let us note that this leaves us with an action of $G$ on the space
$\mc{A}^{(a,b)}$ of $\iG$ invariants.
The latter will become important later on. 
\smallskip

We can realize the $G$-module $\mc{A}^{(a,b)}$ explicitly in terms 
of $\iG$-equivariant functions on the group $G$, 
\begin{equation}
\label{eq:BraneAlgebra}
  \mc{A}^{(a,b)}
  \ \cong\ \Bigl\{ \tA \in \mc{F}^{(a,b)} \Bigl|
   \tA(gh)=\tR_b(h)^{-1}\tA(g)\tR_a(h)\text{ for }h\in \iG\Bigr\}.
\end{equation}
When the two involved representations are trivial, i.e.\ $a=b=0$,
elements of $\mc{A}^{(a,b)}$ are simply invariant under right 
translations with respect to $\iG\subset G$.
\smallskip

There exists more structure on the spaces $\mc{A}^{(a,b)}$ if 
$a=b$. In fact, $\mc{A}^{(a)} = \mc{A}^{(a,a)}$ inherits an 
associative product from the pointwise multiplication of elements 
in $\mc{F}^{(a,a)}$. This turns the subspace $\mc{A}^{(a)}$   
of $\iG$-invariants into an associative matrix algebra. 
\smallskip 

The constructions we have outlined so far may easily be generalized
to arbitrary superpositions of D-branes. To this end we replace 
the irreducible representations~$V_a,V_b$ in~\eqref{eq:DBraneModule} 
by reducible ones. Let $V_{Q}$ be such a reducible representation, 
i.e. $V_{Q}\cong\oplus Q^a V_a$. It represents a superposition 
of $\sum Q^a$ D-branes in which $Q^a$ branes of type $a\in
\reps_{\iG}\cong\treps$ are placed on top of each other. 
Strings ending on such a brane configuration~$Q$ give rise to 
an algebra $\mc{A}^{(Q)}=\mc{A}^{(Q,Q)}$ analogous to~\eqref{eq:DBraneModule}. 
For a stack of $N$~identical branes of type~$a\in\reps_{\iG}
\cong\treps$, the constructions specialize and produce the 
typical Chan-Paton factors, 
\begin{equation}
  \label{eq:BraneStackAlgebra}
  \mc{A}^{N(a)}
  \ \cong\ \Inv_{\iG}\Bigl(\mc{F}^{(a,a)}\Bigr) \otimes \Mat(N)
  \ \ . 
\end{equation}
Obviously, the left translation of the group $G$ turns this into 
a $G$-module with trivial action of $G$ on~$\Mat(N)$.
\medskip

This concludes the formulation of our proposal for the algebra 
of ``functions on twisted D-branes''. There exist two different
kinds of evidence which we can use to motivate and support our 
claim. Let us begin with a simple semiclassical argument. It 
is not difficult to see that a twisted conjugacy class ``close to'' 
the twisted conjugacy class of the group unit can be represented 
in the form $C^\omega = G \times_{\iG} C'$. Here, $C'$ is a 
conjugacy class of $\iG$ ``close to'' the group unit and $\iG$ 
acts on $G$ by multiplications on the right. In other words, 
$C^\omega$ can be considered as a bundle over $G/\iG$ with 
fiber $C'$. In the $k\to \infty$ limit, we keep $C'$ small by
rescaling its radius. As discussed in \cite{Alekseev:1999bs}, 
$C'$ then turns into a co-adjoint orbit. At the same time, the 
volume of $G/\iG$ grows with $k$ so that the corresponding Poisson 
bi-vector scales down. Hence, the $G/\iG$ part of $C^\omega$ 
becomes classical. After quantization, we obtain a bundle with 
non-commutative fibers. While the co-adjoint orbits turn 
into $\Hom(V_a,V_a)$, the base $G/\iG$ stays classical in 
agreement with our claim for $a = b$.  
\smallskip

More substantial support comes from the exact CFT results 
described in the last section. In particular, we shall 
confront the formula~\eqref{eq:BraneAlgebra} with the 
CFT-spectrum of boundary fields~\eqref{eq:CFTSpectrum}. 
Before we carry out the details, let us note that the  
sewing constraints~\eqref{eq:SewingConstraints} are 
automatically satisfied by our construction if we manage
to show that the spectra match. In fact, associativity is
manifest in our proposal and it is the only content of the 
sewing constraints when we send the level $k$ to infinity. 
\smallskip

Hence, it remains to discuss the spectrum of open strings.
The CFT description provides an expression eq.~\eqref{eq:CFTSpectrum}
for the spectrum of strings stretching between D-branes of 
type~$\alpha,\beta\in\tareps$ in terms of characters of
$\asg_k$ which explicitly shows the $G$-module structure 
of the space of ground states that emerges in the limit 
$k\rightarrow\infty$. We claim that in this limit, the $G$-module 
of ground states is isomorphic to the $G$-module $\mc{A}^{(a,b)}$ 
where $a=\Iso^{-1}(\alpha)$ and $b=\Iso^{-1}(\beta)$ 
are the pre-images of the boundary labels $\a$ and $\b$ under the 
map $\Iso: P^+_{\iG} \rightarrow \tareps$ (see above). We 
will prove this by decomposing  $\mc{A}^{(a,b)}$ into irreducibles. 
To do so, let us note that there is a canonical isomorphism $\Hom(V,W)
\cong V^\ast\otimes W$. Furthermore, we may apply the Peter-Weyl 
theorem to decompose the algebra $\mc{F}(G)$ with respect to the 
regular action of $G\times G$ into
\begin{equation*}
  \mc{F}(G)\ \cong \ {\bigoplus}_{A}\, U_A^\ast\otimes U_A
\end{equation*}
where~$A$ runs over all irreducible representations of~$G$
and the two factors of~$G\times G$ act on the two vector spaces
$U_A^\ast,U_A$, respectively. To make contact with our definition of 
$\mc{A}^{(a,b)}$, we have to restrict the right regular $G$~action 
to the subgroup~$\iG$, which leaves us with the 
$G \times \iG$-module 
\begin{equation*}
  \mc{F}(G)\ \cong \ {\bigoplus}_{A,c}\, {b_A}^c\, U_A^\ast\otimes V_c\ \ .
\end{equation*}
The numbers~${b_A}^c\in\Natural_0$ are the so-called {\em branching 
coefficients} which count the multiplicity of the $\iG$-module~$V_c$ 
in~$U_A$. Combining these remarks we arrive at 
\begin{equation*}
  \mc{F}^{(a,b)}
  \ \cong\  \bigoplus_{A,c}\, {b_A}^c\, U_A^\ast
   \otimes V_c\otimes V_a^\ast\otimes V_b \ \ . 
\end{equation*}
It remains now to find the invariants under the $\iG$-action. 
Note that $\iG$ acts on the last three tensor factors. The 
number of invariants in the triple tensor product of irreducible 
representations is simply given by the fusion rules ${N_{cb}}^a$ 
of $\iG$. Hence, as a $G$-module, we have shown that  
\begin{equation*}
  \mc{A}^{(a,b)}
  \ \cong \ \bigoplus_{A,c}\, {b_A}^c{N_{cb}}^a\, U_A^\ast
\end{equation*}
This decomposition is now to be compared with the formula 
\eqref{eq:CFTSpectrum} for the CFT partition functions.
A careful analysis shows that both decompositions agree 
in the limit~$k\to\infty$ provided one uses the appropriate 
identification map $\Iso$, i.e.\ 
\begin{equation} \label{magform} 
  {\bigl(n_A^\omega\bigr)_\alpha}^\beta
  \ = \ \sum_{c}\, {b_{A}}^c\, {N_{ca}}^b
\end{equation} 
where the indices are related by $\alpha=\Iso(a)$ and
$\beta=\Iso(b)$.
The proof is quite technical and not relevant for the further 
developments in this paper. It can be found in Appendix 
\ref{sc:Correspondence}. Let us only mention at this 
place that our proof takes some benefit of results found
in~\cite{Quella:2001wh,Quella:2002wi}.
\smallskip

As a simple cross-check we consider the case of trivial automorphism
$\omega=\id$ where we can make contact to well known results (see  
\cite{Alekseev:1999bs}). First we observe that the construction 
above simplifies considerably since $\iG \cong G$. This implies 
that all the lower case labels can be replaced by capital letters.
In particular, the boundary conditions are now labeled by 
representations of~$G$ itself which is a well established fact. 
The corresponding $G$-module structure is now given by
\begin{equation*}
  \mc{A}^{(A,B)}
  \ \cong \ \bigoplus_C \, {N_{A^+B}}^{C} \, U_{C}\ \ .
\end{equation*}
This is in complete agreement with the known CFT results in
{\em Cardy's case}~\cite{Cardy:1989ir}.%
\medskip%

We close this section with an interesting side-remark. Since we 
were interested in the analysis of twisted branes, our presentation 
has focused on the subgroup $\iG$ in $G$. The right hand side of 
our central formula \eqref{magform}, however, gives rise
to an infinite-dimensional analogue of a NIM-rep\footnote{{\bf N}on-negative
{\bf I}nteger valued {\bf M}atrix {\bf Rep}resentation.} for the fusion
algebra\footnote{The latter is obtained from the fusion algebra of the WZW
model in the limit $k\rightarrow \infty$.} of the Lie group $G$ and any choice
of a subgroup $H \embin G$.
This has been discussed extensively in \cite{Quella:2001wh} where also
the connection of the $k\to\infty$ limit of NIM-reps for twisted boundary
conditions in WZW models to these NIM-reps has been established.
Recently, new explicit expressions for NIM-reps of twisted D-branes have been
proposed in \cite{Petkova:2002yj,Gaberdiel:2002qa} for finite level $k$.
They bear some similarity with our formula \eqref{magform} but generically
do not involve branching coefficients of the invariant subgroup $\iG$.
Where they do, i.e. for the $A_n$-series in \cite{Petkova:2002yj} as well as
the $D_n$-series and the $A_{2n}$-series in \cite{Gaberdiel:2002qa},
they reduce to relation \eqref{magform} in the limit $k\to\infty$.
It seems to us that so far a systematic understanding of the proper
choice for the relevant subgroup is only available at infinite level.

% -----------------------------------------------------------------------
% -----------------------------------------------------------------------
\subsection{The non-commutative gauge theory}

Equipped with the exact solution of the boundary WZW~model in the
limit~$k\rightarrow\infty$, we are finally prepared to calculate 
the low-energy effective action for massless open string modes. 
Compared to the case of D-branes in flat space with background 
$B$-field which leads to a Yang-Mills theory on a non-commutative 
space~\cite{Seiberg:1999vs}, there are two important changes in 
the computation. First, the non-commutative Moyal-Weyl product 
gets replaced by the product of $\iG$-equivariant matrix valued 
functions~\eqref{eq:BraneStackAlgebra} described in the previous
subsection. Moreover, there appears a new term ${f_{\mu\nu}}^{\sigma} 
J_\sigma$ in the operator product expansion of the currents. This 
term leads to an extra contribution of the form  $f_{\mu\nu\sigma}
\tA^\mu\tA^\nu\tA^\sigma$ in the scattering amplitude of three 
massless open string modes. Consequently, the resulting effective 
action is not only given by Yang-Mills theory on a non-commutative space 
but also involves a Chern-Simons like term.% 
\smallskip%

For~$N$ branes of type~$a\in\reps_{\iG}\cong\treps$ on top of each 
other, the fields~$\tA^\mu(g)$ are elements of~$\mc{A}^{N(a)}$, i.e.
they are functions on~$G$ with values in $\End(V_a)\otimes \Mat(N)$ 
and equivariance property as formulated in eq.~\eqref{eq:BraneAlgebra}.
We denote the dimension of~$V_a$ by~$d_a$.
The results of the computation~\cite{Alekseev:2000fd} may easily be 
transferred to the new situation and can be summarized in the 
following action
\begin{equation}
  \label{effact}
  \cS_{N(a)}\ =\  \cS_{{\rm YM}} + \cS_{{\rm CS}}\
     =\frac{\pi^2}{k^2d_aN}\
      \Biggl(\frac{1}{4}\ \int\tr \left( \tF_{\mu\nu} \ \tF^{\mu\nu} \right) 
             - \frac{i}{2}\  \int\tr \left( \tf^{\mu\nu\sigma}\; \tCS_{\mu\nu\sigma}
      \right)\Biggr)
\end{equation}
 where we defined the ``curvature form'' $\tF_{\mu\nu}$ by the expression     
\begin{equation}
  \label{fieldstr}
  \tF_{\mu\nu}(\tA)\, = \,   
   i\, \tL_\mu \tA_\nu - i\, \tL_\nu \tA_\mu + i \,[ \tA_\mu \astk \tA_\nu] 
   +  \tf_{\mu\nu\sigma} \tA^\sigma
\end{equation}
and a non-commutative analogue of the Chern-Simons form by 
\begin{equation}
  \label{CSform}
  \tCS_{\mu\nu\sigma}(\tA)\, = \, \tL_\mu \tA_\nu \, \tA_\sigma 
                   + \frac{1}{3}\; \tA_\mu \, [ \tA_\nu \astk \tA_\sigma]
                   - \frac{i}{2}\; \tf_{\mu\nu\rho}\; \tA^\rho \, \tA_\sigma \ \ .
\end{equation}
Gauge invariance of \eqref{effact} under  the infinitesimal 
gauge transformations
\begin{equation}
\label{eq:GaugeTrafos}
 \delta\tA_\mu  \ = \ i\,\tL_\mu \Lambda \ +\  i\, [\, \tA_\mu\, ,\, 
    \Lambda\, ]  \ \ \ \text{ for } \ \ \ \Lambda \in 
    \mc{A}^{N(a)} 
\end{equation}
follows by straightforward computation. The operator $\tL$ is the 
usual Lie derivative as defined in \eqref{eq:LieDerivative} below.
Note that the ``mass term'' 
in the Chern-Simons form~\eqref{CSform} guarantees the gauge 
invariance of $\cS_{{\rm CS}}$. On the other hand, the effective 
action~\eqref{effact} is the unique combination of $\cS_{\rm YM}$ 
and $\cS_{\rm CS}$ in which mass terms cancel. 
\smallskip

In contrast to earlier work, the trace is now normalized by 
$\tr(\id)=d_aN$. Moreover, we use conventions in which only pure 
Lie algebraic quantities appear. The remaining part of this 
section is devoted to presenting these conventions which 
follow~\cite{Fuchs:1995}. Indices are raised and lowered 
using the Killing form
\begin{equation}
  \label{eq:KillingForm}
  \kappa(x,y)\ =\ \tr(\ad_x\circ\ad_y)\hspace{2cm}
  \kappa^{\mu\nu}\ =\ \frac{1}{I_\theta}\:\kappa(\:\tT^\mu,\tT^\nu\:)
   \ =\ \frac{1}{I_\theta}\:f^{\mu\rho\sigma}\:{f^{\nu}}_{\rho\sigma}
 \ \ .
\end{equation}
Here, $I_B$ denotes the Dynkin index of a representation~$B\in\reps$
and $\tT^\mu$ are the generators of $\mf{g}$ satisfying
\begin{equation} 
\label{eq:LieAlgebra} 
[\tT^\mu,\tT^\nu] \ =\ i{f^{\mu\nu}}_\sigma\tT^\sigma\ \ .  
\end{equation}   
For the adjoint representation~$\theta$ of~$\mf{g}$ 
entering~\eqref{eq:KillingForm}, the Dynkin index may also be 
expressed by the quadratic Casimir and by the dual Coxeter number 
according to $I_\theta=C_\theta=2g^\vee$. The quadratic Casimir 
and the index of an  arbitrary representation~$B\in\reps$ with
dimension~$d_B$ are given by
\begin{equation}
  \label{eq:QuadraticCasimir}
  C_B \ =\ (B,B+2\rho)
  \qquad\qquad\text{ and }\qquad\qquad I_B\ =\ 
  \frac{d_B}{\dim\mf{g}}\, C_B\ \ .
\end{equation}
It will be of particular importance for us to evaluate 
the quadratic Casimir in a given representation~$B \in 
\reps$ according to 
\begin{equation}
  \label{eq:RepQuadraticCasimir}
  \tr\:\tR_B(\tT^\mu)\tR_B(\tT^\nu)\:=\:I_B\:\kappa^{\mu\nu}\ \ .
\end{equation}
Finally, let us note that the $G$-action on the 
algebra~$\mc{A}^{N(a)}$ allows us to define the derivatives
\begin{equation}
  \label{eq:LieDerivative}
  \tL^\mu\tA(g)
  \ =\ \frac{1}{i}
   \frac{\text{d}}{\text{d}t}\Bigl(\tA(e^{-it\tT^\mu}g)\Bigr)\Bigr|_{t=0}
\end{equation}
for~$A\in\mc{A}^{N(a)}$. These Lie derivatives appear in the 
construction of the action above and they satisfy the same 
Lie algebra relations as in eq.~\eqref{eq:LieAlgebra}. 

For a calculation in the next section we will also need derivatives on
functions $\tK: G\to \Mat(N)$ which are defined via the group action on the
argument from the right,
\begin{equation}
  \label{eq:RightDerivative}
  \tL^{\!\tR\mu}\tK(g)
  \ =\ \frac{1}{i}
   \frac{\text{d}}{\text{d}t}\Bigl(\tK\bigl(ge^{it\tT^\mu}\bigr)\Bigr)\Bigr|_{t=0}\ \ .
\end{equation}
The connection between the different derivatives
\eqref{eq:LieDerivative},\eqref{eq:RightDerivative} is given by the adjoint
representation of $G$,
\begin{equation}
\tL_{\mu}\tK(g)\ =\ - \text{Ad}(g^{-1}){_{\mu}}^{\nu} \ 
\tL^{\!\tR}_{\nu} \tK(g)\ \ .
\end{equation}

% -----------------------------------------------------------------------
% -----------------------------------------------------------------------
% -----------------------------------------------------------------------
\section{Condensates of branes on group manifolds}

The study of classical solutions of the non-commutative gauge 
theories constructed in the previous sections provides insights
into the dynamics of twisted branes on group manifolds. In the 
first subsection we shall exhibit a rather large class of 
solutions. The symmetry preserving solutions are then interpreted
geometrically in the second subsection. In particular, we shall 
see that all twisted D-branes appear as bound states from stacks 
of a distinguished elementary twisted brane.   

\subsection{\label{sc:Solutions}Classical solutions}

A simple  variation of the action~\eqref{effact} with respect to 
the gauge fields allows us to derive the following equations of 
motion, 
\begin{equation}
\label{eom}
\tL^{\mu}\tF_{\mu\nu}+[\tA^{\mu},\tF_{\mu\nu}]\ =\ 0 \ \ ,
\end{equation}
which express that the curvature has to be covariantly constant.
Note that the space of gauge fields, and hence the equations 
of motions, depends on the brane configuration we are looking at.
\smallskip 

Solutions to the equations \eqref{eom} describe condensation processes 
on a brane configuration $Q$ which drive the whole system into 
another configuration $Q'$. 
To identify the latter, we have two different types of information
at our disposal. On the one hand, we can compare the tension of 
D-branes in the final configuration $Q^\prime$ with the value of the 
action $\cS_Q(\tA)$  at the classical solution $\tA(g)$. On the other 
hand, we can look at fluctuations around the chosen solution and 
compare their dynamics with the low energy effective theory 
$\cS_{Q^\prime}$ of the brane configuration $Q^\prime$. In formulas, 
this means that 
\begin{equation}\label{fluctuation}
\cS_{Q} (\tA+\delta \tA)\ \overset{!}{=}\  \cS_{Q} (\tA) + 
\cS_{Q^\prime} (\delta \tA) \ \ \ \text{ with } \ \ \
\cS_Q(\tA) \ \overset{!}{=} \ \ln \frac{g_{Q^\prime}}{g_Q} 
\ \ . 
\end{equation}
The second requirement expresses the comparison of tensions in 
terms of the g-factors of the involved  conformal field theories
(see e.g.\ \cite{Alekseev:2000fd} for more details). All equalities  
must hold to the order in $(1/k)$ that we used when constructing 
the effective actions. We say that the brane configuration $Q$ decays
into $Q'$ if $Q'$ has lower mass, i.e.\ whenever $g_{Q'}<g_{Q}$. 

In terms of the world-sheet description, each classical solution of the
effective action is linked to a conformal boundary perturbation in the
CFT of the brane configuration $Q$. Adding the corresponding boundary
terms to the original theory causes the boundary condition to change
so that we end up with the boundary conformal field theory of another
brane configuration $Q'$. Recall, however, that all these statements
only apply to a limiting regime in which the level $k$ is sent to infinity.
\smallskip

We are especially interested in processes that connect maximally
symmetric configurations of D-branes, i.e.\ those configurations 
for which the spectrum decomposes into representations of~$\asg_k$. 
For such configurations we know the low energy effective actions 
so that we are able to compare them with the dynamics of fluctuations. 
We shall argue that solutions possessing a $G$-symmetric fluctuation 
spectrum are associated with gauge fields $\tA(g)$ whose curvature 
is proportional to the identity, or equivalently 
\begin{equation}\label{symmsol}
[\tF_{\mu\nu} (\tA),\:\cdot\:]\ =\ 0 \ \ .
\end{equation}
The $G$-module structure of the fluctuation spectrum around the 
solution can be characterized by the simple rule  
\begin{equation}\label{derivatives}
\tL^{Q^\prime}_{\mu}\ := \ \tL^{Q}_{\mu}+[\tA_{\mu},\:\cdot\:]\ \ .
\end{equation}
Our notation suggests that we want to think of $\tL^{Q^\prime}_\mu$ 
as derivative associated with some new brane configuration $Q^\prime$. 
Note, however, that by construction it acts on on the space $\mc{A}^
{(Q)}$ of gauge fields on $Q$. To check our claims, we must show that 
the derivatives \eqref{derivatives} satisfy the Lie algebra 
commutation relations and that the action for the fluctuation 
fields involves these new derivatives rather than $\tL^Q$. With 
the Ansatz~\eqref{derivatives} it is easy to show that 
\begin{equation*}
[\tL^{Q^\prime}_{\mu},\tL^{Q^\prime}_{\nu}] \ =\ 
i\,{f_{\mu\nu}}^{\sigma}\tL^{Q^\prime}_{\sigma}-i\,[\tF_{\mu\nu}
(\tA),\:\cdot\:]\ \ . 
\end{equation*}
Furthermore, the expansion of the action around the solution 
$\tA(g)$ reads 
\begin{equation*}
\cS_{Q} (\tA+\delta \tA) \ =\ \cS_{Q} (\tA)+\cS_{Q^\prime} 
(\delta \tA)+\frac{i}{2}\,\tr\bigl([\tF_{\mu\nu} (\tA),\delta 
\tA^{\mu}]\delta \tA^{\nu}
\bigr) \ \ .
\end{equation*}
In a slight abuse of notation we have denoted the action 
for the fluctuation fields $\delta \tA_\mu \in \mc{A}^{(Q)}$ 
by $S_{Q^\prime}$. For the moment, this only means that 
all derivatives $\tL^Q$ are replaced by $\tL^{Q^\prime}$.   
As one reads off from the previous equation, gauge fields 
satisfying \eqref{symmsol} do indeed lead to a $G$-symmetric 
fluctuation spectrum. A more careful investigation shows 
that the converse is also true, which is to say, any solution which
produces a symmetric fluctuation spectrum is necessarily of 
the form we have described. The crucial observation in proving this
statement is that the relation \eqref{derivatives} 
between the derivatives is dictated by comparing third-order 
terms in \eqref{fluctuation}.
\medskip

Before we dive into the discussion of these symmetric solutions, let
us briefly mention some solutions of~\eqref{eom} which break the 
$G$-symmetry. One particular type of non-symmetric solution can be
obtained for each choice of a subalgebra~$\mf{h}$ in $\mf{g}$. We 
label the subalgebra generators by $i,j,\ldots$, the directions 
orthogonal to~$\mf{h}$ by $r,s,\ldots$ Now let~$\tR$ be a $d_{\tR}$-%
dimensional representation of $\mf{h}$. Then the equations of motion 
for a stack of~$d_{\tR}$ identical branes possess a constant solution 
of the form $\tA^{i}=\id \otimes \tR(\tT^i)$ and  $\tA^{r}=0$, where
$\id$ is the identity matrix in the space-time degrees of freedom. 
The curvature of these solutions is given by $\tF_{ij}=0$, $\tF_{ir}=0$
and $\tF_{rs}={f_{rs}}^{i}\tR(\tT_i)$. Obviously, this solution does 
not satisfy~\eqref{symmsol} and hence it breaks the $G$-symmetry. One
can identify solutions of this type with some of the symmetry 
breaking boundary theories that were obtained in \cite{Quella:2002ct}. The
associated boundary states preserve the affine Lie algebra of $H$ 
along with the coset chiral algebra for $G/H$.  
\medskip

There is another more trivial class of constant solutions which break 
the $G$-symmetry. In fact, any set of commuting constant matrices~$\tA_{\mu}$ 
is  a solution to~\eqref{eom}. The corresponding curvature is given 
by $\tF_{\mu\nu}={f_{\mu\nu}}^{\sigma}\tA_{\sigma}$, so that in general
$[\tF_{\mu\nu},\:\cdot\:]\neq 0$. The solutions have vanishing action 
and allow for an easy geometric interpretation~\cite{Recknagel:1998ih,
Alekseev:2000fd}:  
they describe translations of the branes in the group manifold. It is 
not surprising that these translations break the symmetry, unless the 
whole stack is translated at once. In the latter case, the fields $\tA$ 
and thus $\tF$ are multiples of the identity as it should be the case 
for a symmetric solution. Let us observe that uniform translations on the 
group manifold are a rather trivial operation, and we are only interested
in analyzing solutions ``up to translations''. One can show that it suffices 
to study solutions with traceless fields $\tA(g)$. From these it is possible
to generate any other solution by a uniform translation.
\medskip 

After this digression we come back to the study of symmetric solutions. 
It follows from the last remark in the previous paragraph that there is
no loss in restricting ourselves to symmetric solutions with vanishing 
curvature $\tF=0$. Let us now start with a stack of~$N$ branes 
corresponding to the trivial representation $0\in P_{\iG}^+$
of $\iG$. A whole class of solutions of~\eqref{eom} with~$\tF=0$ can 
be constructed out of a function $\tK:G\to\GL(N)$ satisfying
\begin{equation}
  \label{solution}
  \begin{split}
  \tK(gh)&\ =\ \tK(g)\tR(h)\qquad\text{ for all }\qquad 
   h\in\iG\qquad\text{ by }\\[2mm] 
  \tA_{\mu}(g)&\ :=\ -(\tL_{\mu}\tK)(g)\tK(g)^{-1}
  \ =\ \tK(g)(\tL_{\mu}\tK^{-1}) (g)
  \end{split}
\end{equation}
with~$\tR$ being an $N$-dimensional representation of~$\iG$.
The first property guarantees the necessary invariance property
$\tA(gh)=\tA(g)$ of the physical fields~$\tA(g)$. Moreover, the 
curvature of these gauge fields vanishes because 
\begin{eqnarray*}
  \tF_{\mu\nu}
  &=& \, i\,\tL_{\mu}\tA_{\nu}-i\tL_{\nu}\tA_{\mu} + i\,[\tA_{\mu},\tA_{\nu}]
     + {f_{\mu\nu}}^{\sigma}\tA_{\sigma}\\[2mm]
  &=& -i\,\tL_{\mu}\bigl(\tL_{\nu}\tK \, \tK^{-1}\bigr)
      +i\,\tL_{\nu} \bigl(\tL_{\mu}\tK\,\tK^{-1}\bigr)\\[1mm]
   && \ \ \ \ -i\,\bigl(\tL_\mu\tK\,\tL_\nu\tK^{-1}-\tL_\nu\tK\,\tL_\mu\tK^{-1}\bigr)
     - {f_{\mu\nu}}^{\sigma}\tL_{\sigma}\tK\,\tK^{-1}\\[2mm] 
  &=& -i\,[\tL_{\mu},\tL_{\nu}]\tK\,\tK^{-1}-{f_{\mu\nu}}^{\sigma}
  \tL_{\sigma}\tK\,\tK^{-1}\ = \ 0  \ \ .
\end{eqnarray*}
We should mention at this point that~$\tF=0$ is precisely the 
integrability condition of the linear system $\tL^{\mu}\tK=-\tA^{\mu}
\tK$ so that any solution to $\tF = 0$ is of the form~\eqref{solution}.
The covariance law $\tK(gh)=\tK(g)\tR(h)$ for $h\in\iG$ is then
necessary to ensure the invariance property~$\tA(gh)=\tA(g)$
for~$h\in\iG$. Consistency implies that~$\tR(h)$ satisfies the 
representation property. 
\smallskip 

Our formula \eqref{solution} for symmetric solutions leaves us 
with the problem to find a $\GL(N,\mathbb{C})$ valued function 
satisfying the covariance condition $\tK (gh)=\tK(g)\tR(h)$ 
for $h \in \iG$. As we shall argue now, such a function does 
always exist. 
\smallskip

To begin with, let us rephrase the existence of $\tK$ in the 
language of vector bundles. For a given representation $\tR$, 
we consider the vector bundle $E\to G/\iG$. Here $E$ is the 
associated bundle $G\times_{\iG}V^\Complex_\tR$ which consists 
of pairs $(g,v)$ along with the identification $(g,v)\sim
\bigl(gh,v\,\tR(h)\bigr)$. Suppose now that we are given the 
function $\tK: G\to\GL(N)$. By construction, the $N$ row-vectors 
$k_i(g)$ are linearly independent and satisfy $k_i(gh) =k_i(g)
\tR(h)$. In other words, they provide a linearly independent 
set of global sections of~$E$, i.e.\ a global trivialization. 
The converse is also true. Hence, the existence of $\tK$ is 
equivalent to the triviality of $E$. 
\smallskip

It is well known that the bundle $G \times_{\iG} V_\tR$ is 
trivial if the representation $\tR$ of $\iG$ can 
be obtained by restricting a representation of $G$. This 
condition is too strong for our purposes. Note, 
however, that we construct our bundle $E$ with the 
complexified space $V_\tR^\Complex$ and this makes a huge 
difference. In fact, $E$ turns out to be trivial for all 
representations $\tR$. 
\smallskip

The proof proceeds in two steps. First one shows that all the 
bundles $E$ are stably equivalent to a trivial bundle. Since 
stable equivalence is preserved under tensoring representations
of $\iG$, one can concentrate on the fundamental representations
and hence reduce the problem to a finite number of checks which 
have to be performed case-by-case. Passing from stable equivalence 
to isomorphism is possible if the rank of the bundle is sufficiently 
large compared to the dimension of the base manifold. Again, this 
has to be decided by going through all the cases separately. The 
details of this proof can be found in Appendix B.   
\smallskip 

After these remarks on the existence of $\tK$, let us briefly address 
the issue of uniqueness. On the face of it, formula \eqref{solution} 
seems to provide a very large 
set of solutions with many continuous parameters. For a given 
representation $\tR $, however, all these solutions are gauge 
equivalent. To see this, let $\tK ,\tA $ and $\tK',\tA'$ be 
two such solutions with a covariance law involving the same $\tR$.  
Then we find 
\begin{equation}
\tA_{\mu}'\ =\ U\tA_{\mu }U^{-1} - (\tL _{\mu }U)U^{-1}
\end{equation}
with $U=\tK'\,\tK^{-1}$, i.e.\ $\tA $ and $\tA'$ are related by a 
gauge transformation $U$ (note that this requires invariance of our
theory under global gauge transformations which is granted as long
as we do not consider the automorphism of order $3$).
In conclusion, our discussion in this 
subsection provides for each $\tR$ a unique solution with vanishing
curvature.
\medskip

The case of~$\tK\equiv\tR$ being a representation of~$G$ can be 
considered as a special example of our general construction as 
it may be restricted to a (reducible) representation of~$\iG$. 
The corresponding field~$\tA(g)$ turns out to be constant and 
it is given by 
\begin{equation*}
  \tA^{\mu} \ =\ \tR(\tT^{\mu})
\end{equation*}
where~$\tT^{\mu}$ are generators of the Lie algebra~$\mf{g}$. 
It is easy to see that all constant solutions to~$\tF=0$ have this 
form. This class of solutions and the corresponding brane processes 
have already been discussed in~\cite{Fredenhagen:2000ei}. In the 
case of untwisted branes, these solutions form a complete set of 
solutions for a stack of branes of type~$0\in P_{G}^+$. For twisted 
branes, however, we just presented new non-constant solutions which 
give rise to new brane processes.

% -----------------------------------------------------------------------
% -----------------------------------------------------------------------
\subsection{\label{sc:Interpretation}Interpretation of the solutions}

In the last section we presented a large number of new stationary points 
for the action of~$N=d_a$ branes of type~$0\in P_{\iG}^+\cong\treps$. 
Here,~$d_a$ is the dimension of an irreducible representation~$\tR_a$ 
of~$\iG$. We shall present evidence that such solutions correspond to 
processes of the type
\begin{equation}
  \label{eq:ConjectureDynamic}
  \text{Stack of }d_a\text{ branes of type }0\in P_{\iG}^+
  \quad\longrightarrow\quad\text{ single brane of type }a\in P_{\iG}^+
\ \ . 
\end{equation}
Any twisted D-brane can be obtained as a condensate of a stack of
elementary branes of type $0\in P_{\iG}^+$. This implies in particular 
that any two configurations whose corresponding representations of 
$\iG$ have the same dimension can be related by a process. A similar 
observation has been made for untwisted
branes~\cite{Alekseev:2000fd, Fredenhagen:2000ei} and the new
processes might provide further insights to the contribution 
of twisted D-branes to the twisted K-groups which seem to 
describe the D-brane charges on group manifolds
(see~\cite{Bouwknegt:2000qt,Fredenhagen:2000ei,Maldacena:2001xj}).
\smallskip

According to eq.~\eqref{fluctuation}, there are essentially two 
checks that we must perform in order to test the 
conjecture~\eqref{eq:ConjectureDynamic}. We will compare the 
value of the action at the solution to the CFT-prediction 
below and start with an analysis of the fluctuations around 
the solution. 
\smallskip

The general form of the fluctuation spectrum follows from the 
discussion in section~\ref{sc:Solutions} and is summarized in 
eq.~\eqref{fluctuation}.
It is implicitly contained in the $G$-action entering the derivatives
\eqref{derivatives} which we used to compute the
action $\cS_{Q'}(\delta \tA)$ of the fluctuation fields. But as
we have stressed earlier, the fields $\delta \tA$ are elements of 
the algebra $\mc{A}^{d_a(0)}$. Let us be precise and introduce 
a new symbol $\widetilde{\mc{A}}^{d_a(0)}$ for the $G$-module 
$\bigl(\mc{A}^{d_a(0)},\tL_{Q^\prime}\bigr)$. Our aim then is to identify 
the functional $\cS_{Q'}(\delta \tA)$ on $\widetilde{\mc{A}}^{d_a(0)}$ 
with the action functional $\cS_{(a)}$ on $\mc{A}^{(a)}$ which 
governs the dynamics of a brane of type $(a)$. This comes down 
to providing a $G$-module isomorphism $\Phi$ between 
$\widetilde{\mc{A}}^{d_a(0)}$ and $\mc{A}^{(a)}$. In particular, 
the derivatives have to match,
\begin{equation*}
\Phi\bigl(\widetilde{\tL}_\mu^{d_a(0)}\ \delta \tA\bigr)\ =\ \tL_\mu^{(a)}
\ \Phi(\delta \tA)\ .
\end{equation*}
Such an isomorphism can be obtained with the help of the function $\tK$ 
from which we constructed our solution,
\begin{equation*}
\Phi(\delta \tA)(g)\ =\ \tK^{-1} (g)
  \, \delta \tA(g) \, \tK (g) \ \in \mc{A}^{(a)} \ . 
\end{equation*}
One can check that the action of the derivatives $\tL_\mu^{(a)}$ 
on $\mc{A}^{(a)}$ precisely coincides with the action of the shifted
derivatives on $\widetilde{\mc{A}}^{d_a(0)}$,
$$\widetilde{\tL}_\mu^{d_a(0)} \ =\ \tL_\mu^{d_a(0)}+
 [\tA_\mu,\:\cdot\:]\ \overset{\Phi}{\cong}\ \tL_\mu^{(a)}\ \ ,$$ 
where the shift is given by our solution
$\tA_{\mu}=-\tL_{\mu}\tK\:\tK(g)^{-1}$ of the 
equations of motion. Thus we have proved that the theory of fluctuations 
around our solution coincides with $\cS_{(a)}$ as we anticipated in 
eq.~\eqref{eq:ConjectureDynamic}.  
\medskip

Our second check involves a comparison between the value of the action 
at the solution and the g-factors of a CFT description. For technical 
reasons, we restrict ourselves to automorphisms $\omega$ of order~$2$, 
thereby excluding only the case of triality for $D_4$. The existence of
$\omega$ 
implies strong constraints on the form of the structure constants. To 
be specific, by diagonalization of $\omega$, we may choose a basis 
in which only the constants $f_{ijk}, f_{rsi}$ and cyclic permutations 
thereof do not vanish. Here, $i,j,k,\ldots$ denote indices for elements 
in the invariant subalgebra~$\ig$ and $r,s,t,\ldots$ label directions 
orthogonal to~$\ig$. We are now able to compute 
the action using no more than the properties of the solution we have 
specified. For a solution with $\tF =0$, the action reduces to
\[
\cS_{d_a(0)}(\tA) \ = \ \kappa f^{\mu \nu \sigma }\int\tr\:\tA_\mu
\tA_\nu \tA_\sigma 
\]
with $\kappa = i \pi^2/ 6k^2 d_a$. We now express the solution 
\eqref{solution} in terms of the right derivatives
\eqref{eq:RightDerivative}, 
$$
  \tA_{\mu }\ =\ \Ad
   (g^{-1}){_{\mu }}^{\nu }\bigl(\tL^{\!\tR }_{\nu }
 \tK\bigr)\tK^{-1}\ \ . 
$$
Since we have to contract the $\tA_\mu$s with $f^{\mu \nu \sigma }$, 
the adjoint action of $G$ can be dropped to obtain 
\[
\cS_{d_a(0)}(\tA) \ = \ \kappa f^{\mu \nu \sigma }\int\tr\:\tA'_\mu
\tA'_\nu \tA'_\sigma
\]
with $\tA '_{\mu }=\tK^{-1}(\tL ^{\!\tR }_{\mu }\tK)$. Our previous
remarks on the form of the structures constants suggest to split this
formula for the action into two terms 
\begin{equation}
  \label{eq:AuxAction}
  \cS_{d_a(0)}(\tA) \ = \ \kappa f^{ijk}\int\tr\:\tA'_i\tA'_j\tA'_k
   +3 \kappa f^{rsi}\int\tr\:\tA'_r\tA'_s\tA'_i\ \ .
\end{equation}
The first term can be computed 
easily since $\tA'_{i}(g)=\tR_a(\tT_i)$. 
As for the second term, one proceeds by expressing the gauge field
components $\tA'_r(g)$ through $\tK(g)$. Integration by parts and 
use of the anti-symmetry of the structure constants results in 
$$ \cS_{d_a(0)} (\tA) \ = \ \kappa f^{ijk}\int\tr\:
   \tR_a(\tT_i)\tR_a(\tT_j)\tR_a(\tT_k)
   +\frac{3\kappa }{2} 
   f^{rsi}\int\tr\: \tK^{-1}[\tL^{\!\tR }_r,\tL^{\!\tR }_s]\tK\:\tR_a(\tT_i)
\ \ . $$ 
Because of the constraints on the structure constants, the commutator 
of the derivatives lies in the direction of~$\ig$ so that we obtain 
a factor $i{f_{rs}}^j\tR_a(\tT_j)$ within the trace. After a bit of algebra 
using eqs.~\eqref{eq:KillingForm}, \eqref{eq:QuadraticCasimir} 
and~\eqref{eq:RepQuadraticCasimir}, the two terms in the expression for 
the action can be combined into the following result 
\begin{equation}
  \label{eq:EnergyEff}
  \cS_{d_a(0)}(\tA)\ =\ 
-\frac{\pi^2}{12k^2}\frac{C_a}{x_{e}}\Biggl(3C_\theta
-2\frac{C_{\bar{\theta}}}{x_{e}}\Biggr)\ \ .
\end{equation}
The numbers~$C_a,C_\theta,C_{\bar{\theta}}$ are the quadratic Casimirs 
of the representation~$a$ of $\ig$ and of the adjoint representations 
of $\mf{g},\:\ig$, respectively ($\theta,\bar{\theta}$ are the highest 
roots). The constant $x_{e}$ denotes the embedding index
of the embedding $\ig \hookrightarrow \mf{g}$. It appears
due to different normalization of the quadratic 
forms~$\kappa_{\ig}$ and~${\kappa_{\mf{g}}|_{\ig}}$.
It turns out that the value of the action~\eqref{eq:EnergyEff} 
is always negative.
\smallskip

As we recalled before, this result for the value of the action must 
be compared with the difference between two logarithms of the 
g-factors~\cite{Affleck:1991tk} in the CFT-description. For the 
branes $(a)$, the g-factor is given by \cite[eq.\ (4.2)]{Behrend:1999bn}
\begin{equation}
g_{(a)}\ = \frac{S_{0\Iso(a)}^\omega}{\sqrt{S_{00}}}
\end{equation}
where we explicitly used the identification~$\Iso:P_{\iG}^+\to\treps$ 
to emphasize that the second argument of the twisted S~matrix is a 
fractional symmetric weight. This leads to the following 
expression for the logarithm of ratio of g-factors  
\begin{equation}
  \label{eq:Energygfac}
  \ln\frac{g_{(a)}}{d_ag_{(0)}}
  \ = \ \ln\frac{S_{0\Iso(a)}^\omega}{d_aS_{0\Iso (0)}^\omega}
  \ =\ -\frac{\pi^2}{12k^2}\,\frac{\dim\mf{g}}{\dim\ig}\,\frac{C_a}{x_{e}}\,C_\theta\ \ .
\end{equation}
To derive this result one shows first
that the quotient in the argument of the
logarithm is an ordinary character of $\ig$,
\[
  \frac{S^{\omega}_{0\Iso(a)}}{S^{\omega}_{0\Iso (0)}}\ = \
  \chi_a\biggl(-\frac{2\pi i}{k+g^\vee}\frac{1}{x_e}\mc{P}(\rho)\biggr) \ \ .
\]
Using (a generalized
version of) the asymptotic expansion in~\cite[eq.\ (13.175)]{FrancescoCFT}
one then  recovers~\eqref{eq:Energygfac} to lowest order in~$1/k$.
\smallskip

Although it is not obvious, formula~\eqref{eq:Energygfac} agrees with 
the previous expression~\eqref{eq:EnergyEff} for the value of the 
effective action at the solution which was obtained under the 
assumption that~$\omega$ is an order~$2$ automorphism. 
This can be checked with the help of Table~\ref{tb:Overview} and well 
known Lie algebra data (see for example~\cite[p.\ 44]{Fuchs:1995}).
To summarize we have shown (except for the case $G_2\embin D_4$) that
our solutions describe processes of the type~\eqref{eq:ConjectureDynamic}.

% -----------------------------------------------------------------------
% -----------------------------------------------------------------------
% -----------------------------------------------------------------------
\section{Conclusions and Outlook}

In this work we have extended the previous analysis of brane 
dynamics \cite{Alekseev:1999bs,Alekseev:2000fd,Alekseev:2000wg} 
to brane configurations of all maximally symmetric branes on 
group manifolds including the so-called twisted D-branes. In 
particular, we exploited the CFT data to construct the algebra 
of gauge fields on such branes and provided a formula for the 
effective action. Moreover, we found a large number of solutions 
and their geometric interpretation. The new condensation processes 
turn out to be consistent with the charge conservation laws 
formulated in~\cite{Fredenhagen:2000ei} (see also 
\cite{Maldacena:2001xj}). In particular they suggest that the charge 
of an arbitrary $\omega$-twisted D-brane of type~$a\in\reps_{\iG}
\cong\treps$ is given by the dimension of the group 
representation~$a$.
\medskip 

All this analysis is performed in the limiting regime where 
the level $k$ is sent to infinity. It would be interesting to 
investigate how the described condensation processes deform 
when we go to finite values of the level~$k$. For constant
gauge fields, such deformations have been studied in 
\cite{Alekseev:2000jx,Fredenhagen:2000ei} based on the 
``absorption of the boundary spin''-principle (see 
\cite{Affleck:1991by,Affleck:1991iv}). These investigations 
lead to very strong constraints on the structure of the charges 
that are carried by untwisted D-branes. For non-trivial~$\omega$, 
however, constant condensates provide only a small number of 
processes which are difficult to evaluate. The bound state 
formation \eqref{eq:ConjectureDynamic} we found in the last 
section of this paper suggests some obvious extensions that 
might place the analysis of conservation laws for twisted and 
untwisted D-branes on an equal footing. 

Geometrically, the situations with finite~$k$ are associated 
with a non-vanishing NSNS 3-form H-field. As has been argued 
in \cite{Alekseev:1999bs,Cornalba:2001sm,Hayasaka:2001an} 
this might lead to new phenomena in the world-volume geometry.
For branes on group manifolds they seem to be related to 
quantum groups or appropriate modifications thereof 
(see \cite{Alekseev:1999bs,Alekseev:1995rn,Pawelczyk:2002kd}). 
\medskip

Let us also mention that the results we have obtained here can 
be of direct relevance for the study of branes in coset models
that has recently attracted some attention~\cite{Stanciu:1998sk,  
Fredenhagen:2001nc,Maldacena:2001ky,Gawedzki:2001ye,Elitzur:2001qd,
Fredenhagen:2001kw}. In fact, it was shown in \cite{Fredenhagen:2001nc,
Fredenhagen:2001kw} that many results on the dynamics of branes on group 
manifolds descend to cosets through some reduction procedure. The 
main ideas of this reduction are not specific to trivial gluing 
conditions and generalize immediately to twisted branes in coset 
theories. Finally, following the ideas of \cite{Maldacena:2001ky,
Quella:2002ct}, the coset construction is an essential tool to obtain  
symmetry breaking boundary conditions e.g.\ on group manifolds. It 
is therefore tempting to conclude that a combination of all these 
results can provide a rather exhaustive picture of brane dynamics
on group manifolds, even when we go beyond maximally symmetric 
branes.

% -----------------------------------------------------------------------
\subsection*{Acknowledgements}

We would like to thank V.\ Braun, G.\ Moore, I.\ Runkel and 
Ch.\ Schweigert  for their comments and useful discussions. 
The work of T.Q.\ and S.F.\ was supported by the 
Studienstiftung des deutschen Volkes. Part of the research was done 
during a stay of T.Q.\ at the University of Geneva which was supported
by the Swiss National Science Foundation. 

\newpage
% -----------------------------------------------------------------------
% -----------------------------------------------------------------------
% -----------------------------------------------------------------------
\begin{appendix}

% -----------------------------------------------------------------------
% -----------------------------------------------------------------------
% -----------------------------------------------------------------------
\section{\label{sc:Correspondence}Correspondence between boundary 
labels and representations}

  The aim of this section is to provide the correspondence between exactly 
  solvable boundary conditions and representations of the invariant subgroup
  $\iG$. This is needed to relate the results of the more geometric 
  picture to those of CFT. We will proceed in the first subsection by comparing
  the explicit expressions for the NIM-reps~\eqref{annuluscf} to recent results
  for branching coefficients of arbitrary semi-simple Lie
  algebras~\cite{Quella:2001wh}. We are thus able to distinguish a
  certain subalgebra $\sg$ and identify boundary labels with
  representations of $\sg$ such that we get a purely Lie algebra theoretic
  expression for the NIM-reps in the limit $k\to\infty$. In almost all cases
  $\sg$ is given by the invariant subalgebra $\ig$ and we also have
  $P_{\iG}^+=P_{\ig}^+$. The only case where this procedure does not work
  is $\mf{g}=A_{2n}$ where we are lead to the orbit Lie algebra $\sg=C_n$
  and not to the invariant subalgebra $\ig=B_n$.
  It was argued in~\cite{Quella:2001wh}, however, that in
  this specific case a second identification is possible which leads to the
  invariant subalgebra.
  We reserve a second subsection to review
  these results and to discuss the case of $A_{2n}$ in detail. Note that
  results closely related to those presented in this section have also been
  found independently in~\cite{Petkova:2002yj,Gaberdiel:2002qa} for finite $k$
  using different methods. In all cases for which such a finite $k$ extension
  exists, i.e. for the $A_n$-series in \cite{Petkova:2002yj} as well as the
  $D_n$-series and the $A_{2n}$-series
  in \cite{Gaberdiel:2002qa}, our results may also be recovered from the
  existing literature by taking $k$ to infinity.
  Let us emphasize, however, that in the cases of $A_{2n-1}$,
  $D_4$ (triality) and $E_6$ our treatment seems to indicate stronger
  statements, i.e. larger subgroups, for finite $k$ than those proposed
  in \cite{Gaberdiel:2002qa}. The geometrical
  interpretation of these results which was described in the main text and the
  identification of the invariant subgroup as the relevant structure seem to
  be new. The results of this appendix have already been announced in
  \cite{Quella:2001wh} where they were used to derive new
  representations for branching coefficients.

% ------------------------------------------------------------------------
% ------------------------------------------------------------------------
\subsection{The generic correspondence}

In this first subsection we propose an identification of boundary labels
with representations of a distinguished subalgebra $\sg$ of $\mf{g}$ such
that formula~\eqref{magform} is satisfied for the embedding $\sg\embin\mf{g}$
after taking the limit $k\to\infty$.
Neither is obvious that this will work a priori
nor is clear which subalgebra one should take. Starting from certain
assumptions we will first derive a set of consistency relations.
Afterwards we will show that for each $\mf{g}$ there is indeed a unique solution
$\sg$ to these consistency equations and that in most cases it is given by the
invariant subalgebra $\ig$.

Let us begin our discussion with a few general remarks about embeddings
of Lie algebras.  
Any embedding map $\iota:\sg\hookrightarrow\mf{g}$ 
induces a ``projection'' (or rather ``restriction'') in weight space 
$\mc{P}:L_w^{(\mf{g})}\rightarrow L_w^{(\sg)}$. A representation 
is completely defined by its character, i.e.\ by the weight system 
of the given highest weight. Under the projection~$\mc{P}$, this 
weight system is mapped to a sum of weight systems of highest 
weights of the subalgebra. This process may be summarized in the 
decomposition formula $U_A=\oplus_c {b_A}^cV_c$ of modules (both
sides considered as modules of $\sg$) where
we introduced the so called branching 
coefficients~${b_A}^c\in\Natural_0$.
\smallskip

\begin{table}
\centerline{\begin{tabular}{ccccc}
  $\mf{g}$ & order & $\ig$ & orbit Lie 
    algebra~$\breve{\mf{g}}$ & relevant subalgebra~$\sg$ \\\hline
  $A_2$ & 2 & $A_1\,(x_e=4)$ & $A_1$ & $A_1\,(x_e=1)$ \\
  $A_{2n-1}$ & 2 & $C_n$ & $B_n$ & $C_n$ \\
  $A_{2n}$ & 2 & $B_n$ & $C_n$ & $C_n\embin A_{2n-1}$ \\
  $D_4$ & 3 & $G_2$ & $G_2$ & $G_2\embin B_3$ \\
  $D_n$ & 2 & $B_{n-1}$ & $C_{n-1}$ & $B_{n-1}$ \\
  $E_6$ & 2 & $F_4$ & $F_4$ & $F_4$
\end{tabular}}
  \caption{\label{tb:OverviewAppendix}More data
  related to outer automorphisms of simple Lie algebras.}
\end{table}
  
It is now important to specify  which structure our identification 
of boundary labels and representations is supposed to preserve. Let 
$\langle\treps\rangle$ be the integer linear span of the set of 
boundary conditions~$\treps$, i.e.\ the full lattice generated by
elements of~$\treps$. Both the lattice~$\langle\treps\rangle$ and 
the weight lattice~$L_w^{(\sg)}$ permit an action of Weyl groups. 
In the first case this group is given by the symmetric part
$W_{\omega}=\{w\in W_{\mf{g}}|\omega\circ w=w\circ\omega\}$ of 
the Weyl group of~$\mf{g}$ and in the other case it is naturally 
given by~$W_{\sg}$. Furthermore, in both cases we have a projection
$\mc{P}:L_w^{(\mf{g})}\rightarrow L_w^{(\sg)}$ and
$\proj:L_w^{(\mf{g})}\rightarrow\langle\treps\rangle$,
respectively. The latter is given by the projection $\proj
=\frac{1}{N}(1+\omega+\cdots+\omega^{N-1})$ onto the symmetric part
of the weights, $N$ being the order of $\omega$.
As we will see, we have to find an isomorphism $\Iso:L_w^{(\sg)}
\rightarrow \langle\treps\rangle$ between the fractional lattice 
generated by the boundary conditions and the weight lattice of the 
subalgebra which preserves all of these structures. In particular
it should be accompanied with an isomorphism
$\Iso:W_{\sg}\to W_\omega$ of the corresponding Weyl groups.
To summarize, we 
have to find a subalgebra~$\sg$ and a functor-like map~$\Iso$ such
that the diagrams~\eqref{eq:CommDiagram} commute. It turns out
that the answer for both~$\sg$ and~$\Iso$ is unique.
\begin{equation}
  \label{eq:CommDiagram}
\begin{CD}
  L_w^{(\mf{g})} @>{\mc{P}}>> L_w^{(\sg)} &\hspace{2.5cm}&  L_w^{(\sg)} @>w\in W_{\sg}>> L_w^{(\sg)}\\
  @VV{=}V @VV{\Iso}V  @VV{\Iso}V @VV{\Iso}V\\
  L_w^{(\mf{g})} @>{\proj}>> \langle\treps\rangle &\hspace{2.5cm}& \langle\treps\rangle @>\Iso(w)\in W_\omega>>\langle\treps\rangle
\end{CD}
\end{equation}
In the following we restrict ourselves to some non-trivial outer
automorphism~$\omega\neq\id$ since our statements become trivial
otherwise.
\smallskip

Remember that we only consider the limit $k\rightarrow\infty$.
For $\alpha=\Iso(a),\:\beta=\Iso(b)$ we want to
proof the equality
\begin{equation*}
  {\bigl(n_A^\omega\bigr)_{\beta}}^{\alpha}
  \ =\ \sum_c {b_A}^c {N_{cb}}^a
\end{equation*}
where the last expression contains some branching coefficients for
$\sg\hookrightarrow\mf{g}$ and the tensor product coefficients of
the subalgebra. This relation explicitly assumes the existence of the
bijection $\Iso:a\in P_{\sg}^+\leftrightarrow\alpha\in\treps$
conjectured above. Using a result of~\cite{Quella:2002wi}
and a useful identity for 
branching coefficients (which is related to~\cite{Klimyk:1967}, see 
however~\cite{Quella:2001wh} for a recent proof using affine 
Kac-Moody algebra techniques) we may write
\begin{equation}
  \label{eq:NIMrepComparison}
  \begin{split}
  {\bigl(n_A^\omega\bigr)_{\beta}}^{\alpha}
  &\ =\ \sum_{B\in\text{wts}(A),w\in W_{\omega}}\epsilon_{\omega}(w)
  \delta_{\alpha,w(\mathcal{P}_{\omega}B+\beta+\rho_\omega)-\rho_\omega}\\[2mm]
  \sum_c {b_A}^c {N_{cb}}^a
  &\ =\ \sum_{B\in\text{wts}(A),w\in W_{\sg}}\epsilon(w)
  \delta_{a,w(\mathcal{P}B+b+\rho)-\rho}\ \ .
  \end{split}
\end{equation}
The abbreviation $B\in\text{wts}(A)$ means that~$B$ runs over all weights
in the weight system of~$A$. Both expressions are obviously equal to each  
other if the bijection~$\Iso$ is structure preserving, i.e.
\begin{eqnarray*}
  \Iso(\rho)&=&\rho_\omega\\[2mm] 
  \Iso\circ\mc{P}&=&\mc{P}_{\omega}\\[2mm] 
  \Iso(wa)&=&\Iso(w)\Iso(a)\ \ .
\end{eqnarray*}
The last condition already constrains the possible subalgebras to
a large extent. Indeed, the Weyl group~$W_\omega$ can be described as
the Weyl group of the so-called orbit Lie algebra of~$\mf{g}$ with 
respect to the automorphism~$\omega$ (see~\cite{Fuchs:1996zr}).
In some special cases this orbit Lie algebra coincides with the 
invariant subalgebra~$\ig$ while it does not for the whole~$A_{n}$ 
and~$D_n$ series. A survey of these relations can be found in 
Table~\ref{tb:OverviewAppendix} which in part has been taken 
from~\cite{Birke:1999ik}. Note however that the Weyl groups 
of~$B_n$ and~$C_n$ are isomorphic to each other (see 
e.g.~\cite[p.~74]{Fuchs:1995}) which can most easily be seen 
by treating them as abstract Coxeter groups. Thus by imposing
the last constraint we only have to decide which of possibly two
subalgebras - the orbit Lie algebra or the invariant subalgebra -
and which specific kind of embedding we should take. This choice 
is uniquely determined by the other two conditions. In the cases 
of~$A_{2n-1}$ and~$D_n$ the orbit Lie algebra not even is a 
subalgebra so that this possibility is ruled out immediately.
\smallskip

We will show in the most simple example of the Lie algebra~$A_2$
how this procedure works and then state only results for all the 
other cases. Let us consider $\mf{g}=A_2$ with outer automorphism
$\omega(a_1,a_2)=(a_2,a_1)$. The relevant subalgebra is given 
by $\sg=A_1$ and the projection to fractional symmetric weights 
- which describe the boundary conditions of the theory - reads 
$\mc{P}_{\omega}(a_1,a_2)=\frac{1}{2}(a_1+a_2,a_1+a_2)$. There 
are two inequivalent embeddings $A_1\hookrightarrow A_2$ given by
projections $\mc{P}_{x_e}(a_1,a_2)=\sqrt{x_e}(a_1+a_2)$ with embedding 
index~$x_e=1$ and~$x_e=4$ respectively~\cite[p.~534]{FrancescoCFT}.
Imposing the first condition we see that
\begin{equation}
  \Iso\circ\mc{P}_{x_e}(a_1,a_2)
 \  =\ \sqrt{x_e}\:\Iso(a_1+a_2)\ \ .
\end{equation}
This only equals $\mc{P}_{\omega}(a_1,a_2)$ for
\begin{equation}
  \Iso(a)\ =\ \frac{1}{2\sqrt{x_e}}(a,a)\ \ .
\end{equation}
The condition $\Iso(\rho)=\rho_\omega=\frac{1}{2}(1,1)$ forces
us to use the projection with $x_e=1$. One can also check explicitly that 
the Weyl groups correspond to each other. This is the first example
where the relevant subalgebra is not given by the invariant subalgebra
(which has embedding index~$x_e=4$) but by the orbit Lie algebra. The 
same statement holds for the whole $A_{2n}$~series as we will see. As 
mentioned already the results of this section are summarized in 
Table~\ref{tb:OverviewAppendix}.
\smallskip

One can  treat the whole ADE series using a case by case study. Let us
emphasize that we use the labeling conventions for weights which can 
be found in~\cite[p.~540]{FrancescoCFT}. The projections have been 
found using~\cite[p.~57-61]{McKayPateraRand:1990} and the programs 
LiE~\cite{LiE} and SimpLie~\cite{SimpLie}. Note that LiE uses a 
different labeling convention for the weights. For a useful table of 
branching rules see also \cite{McKayPatera:1981}.
\begin{enumerate}
\item The case of~$A_{2n-1}$ is straightforward. The relevant subalgebra
is given by the invariant subalgebra $C_n\embin A_{2n-1}$. This is a
maximal embedding and the identification reads
\[
  \proj
  \ =\ \frac{1}{2}\left(\smat1&&&&&&1\\&\diagdown&&&&\diagup
  \\&&1&&1&&\\&&&2\\&&1&&1&&\\&\diagup&&&&\diagdown\\1&&&&&&1\stam\right)
  \ =\ \frac{1}{2}\left(\smat1&\\&\diagdown\\&&1\\&&&2\\&&1\\&\diagup
  \\1\stam\right)\left(\smat1&&&&1\\&\diagdown&&\diagup\\&&1\stam\right)
  \ =\ \Iso\circ\mc{P}\ \ .
\]

\item The case $A_{2n}$ is exceptional. Here the relevant subalgebra
  is given by the orbit Lie algebra which can be described by the sequence
  of maximal embeddings $C_n\embin A_{2n-1}\embin A_{2n}$
  (for $n=1$ we have $A_1\embin A_2$). The identification reads
\[
  \proj
  \ =\ \frac{1}{2}\left(\smat1&&&&&1\\&\diagdown&&&\diagup
 \\&&1&1&&\\&&1&1&&\\&\diagup&&&\diagdown\\1&&&&&1\stam\right)
  \ =\ \frac{1}{2}\left(\smat1&\\&\diagdown\\&&1\\&&1\\&\diagup
  \\1\stam\right)\left(\smat1&&&&&1\\&\diagdown&&&\diagup\\&&1&1\stam\right)
  \ =\ \Iso\circ\mc{P}\ \ .
\]
  There is a second identification related to the subalgebra
  $B_n\embin A_{2n}$ which will be discussed in the next
  subsection and which is the relevant one for the main part of the paper.

\item The order~$3$ diagram automorphism of~$D_4$ leads to the sequence
  of maximal embeddings~$G_2\embin B_3\embin D_4$ and to the
  identification
\[
  \proj
  \ =\ \frac{1}{3}\left(\smat1&0&1&1\\0&3&0&0\\1&0&1&1\\1&0&1&1\stam\right)
  \ =\ \frac{1}{3}\left(\smat0&1\\3&0\\0&1\\0&1\stam\right)\left(\smat0&1&0&0\\1&0&1&1\stam\right)
  \ =\ \Iso\circ\mc{P}\ \ .
\]

\item For the order~$2$ automorphism of the $D$-series one obtains the
  maximal embedding $B_{n-1}\embin D_{n}$ and
\[
  \proj
  \ =\ \frac{1}{2}\left(\smat2\\&\diagdown\\&&2\\&&&1&1\\&&&1&1\stam\right)
  \ =\ \frac{1}{2}\left(\smat2&\\&\diagdown\\&&2\\&&&1\\&&&1\stam\right)
  \left(\smat1\\&\diagdown&\\&&1\\&&&1&1\stam\right)
  \ =\ \Iso\circ\mc{P}\ \ .
\]

\item Also the last case $E_6$ behaves regular and yields the maximal
  embedding $F_4\embin E_6$ with
\[
  \proj
  =\frac{1}{2}\left(\smat1&0&0&0&1&0\\0&1&0&1&0&0\\0&0&2&0&0&0\\0&1&0&1&0&0\\1&0&0&0&1&0\\0&0&0&0&0&2\stam\right)
  =\frac{1}{2}\left(\smat0&0&0&1\\0&0&1&0\\0&2&0&0\\0&0&1&0\\0&0&0&1\\2&0&0&0\stam\right)\left(\smat0&0&0&0&0&1\\0&0&1&0&0&0\\0&1&0&1&0&0\\1&0&0&0&1&0\stam\right)
  =\Iso\circ\mc{P}.
\]
\end{enumerate}

  These considerations show that in all cases but $A_{2n}$ we may identify
  the boundary labels with representations of the invariant subgroup $\iG$.
  Let us emphasize that it is possible to identify the set of Lie algebra
  representations $P_{\ig}^+$ with the set of group representations $P_{\iG}^+$
  in these cases as the corresponding groups $\iG$ are all simply-connected.
  A detailed discussion of the identification one should use in the
  exceptional case of $A_{2n}$ is postponed to the next subsection.

% ------------------------------------------------------------------------
% ------------------------------------------------------------------------
\subsection{The special case $A_{2n}$}

  In the last section it was shown that under certain natural and
  well-motivated assumptions the relevant subalgebra $\sg$ in the case of
  $A_{2n}$ which describes the boundary labels is uniquely given by the
  orbit Lie algebra $C_n$ and not by the
  invariant subalgebra $B_n$. This contradicts, however, our geometrical
  intuition as we are expecting the invariant subalgebra (or even better
  the invariant subgroup) to be the relevant structure.
  In~\cite{Quella:2001wh}, however, it was argued that one is lead to
  the invariant subalgebra by taking a different inductive limit, i.e.\ an
  identification which involves the level $k$ explicitly. Indeed, in
  writing~\eqref{eq:NIMrepComparison} we already took a very special limit
  implicitly. That taking the limit $k\to\infty$ may have nontrivial effects
  can already be seen from simple current symmetries in fusion rules. In this
  limit all simple current symmetries get lost and it becomes important on
  which ``branch'' of the simple current orbits one sits while taking the
  limit. There is also another point on which we have not been careful enough
  in the last subsection. The geometric picture suggests that we should work
  with representations of the invariant sub{\em group}, not necessarily with
  those of the invariant sub{\em algebra}. These two sets may differ as can
  easily be seen from the familiar example of $SO(3)$ which only allows
  $SU(2)$ representations of integer spin. Note that the automorphism
  $\omega$ in the case of $SU(3)$ is just given by charge conjugation and
  that $SO(3)$ exactly is the invariant subgroup. Similar remarks hold
  for the whole series $SO(2N+1)\subset SU(2N+1)$, i.e.\ the whole
  $A_{2n}$ series. All this
  should be reflected in the new identification in a certain way.

  Let us review the new identification of~\cite{Quella:2001wh} and see
  whether it fits our requirements. The construction only works for even
  values of the level $k$. This fact may be reminiscent of the $D$-series
  modular invariants of $SU(2)$ describing a $SO(3)$ WZW model. Therefore
  we will assume $k$ to be even in what follows. This restriction will not
  be relevant in the limit $k\to\infty$. The labels for the
  twisted boundary conditions in the WZW model based on $A_{2n}$ are given by
  half-integer symmetric weights $\alpha$ of~$A_{2n}$. To be more specific, the
  Dynkin labels have to satisfy the relations $2\alpha_i\in\Natural_0$,
  $\alpha_i=\alpha_{2n+1-i}$ and $\sum_{i=0}^n\alpha_i\leq k/4$. These
  labels may be interpreted as labels of the invariant subalgebra $B_n$
  of $A_{2n}$. The map from weights of $B_n$ to
  the boundary labels is given by~\cite{Quella:2001wh}
\begin{equation}
  \label{eq:NewId}
  \Iso\bigl(a_1,\cdots,a_n\bigr)
  \,=\,\frac{1}{4}\,\Bigl(2a_{n-1}\,,\,\cdots,\,2a_1\,,\,k-2\sum_{i=1}^{n-1}a_i-a_n\,,\,\cdots,\,2a_{n-1}\Bigr)\ \ .
\end{equation}
  Note that this map involves $k$ explicitly and is only well-defined for
  weights whose last Dynkin label $a_n$ is even. This last condition 
  has two interpretations.
  From the group theoretical point of view it restricts to representations
  of the Lie algebra $B_n$ which may be integrated to single-valued
  representations of the group $SO(2n+1)$. From the Lie algebra point of view
  it corresponds to the
  branching selection rule of the embedding $B_n\hookrightarrow A_{2n}$.
  The relevant projection for this embedding reads
  $\mc{P}\bigl(i_1,\cdots,i_{2n}\bigr)
   =\bigl(i_1+i_{2n},i_2+i_{2n-1},\cdots,2(i_n+i_{n+1})\bigr)$.
  
Actually, there is a geometric reason why we should use a
$k$-dependent identification map in the case of $A_{2n}$. 
In the limit $k\to \infty $, the
twisted conjugacy classes in the vicinity of 
the group unit have boundary labels
close to $(0,\dots,0 ,k/4,k/4,0,\dots ,0)$. This can be inferred from
an analysis of brane geometry seen by closed strings \cite{Felder:1999ka}.

Let us summarize these results.
  Using the new identification map~\eqref{eq:NewId} we 
can identify the labels of twisted branes in $SU (2n+1)$
sitting close to the group unit
  with representations of the invariant subgroup $\iG\hookrightarrow G$.
  In particular this identification knows about the fact that
  certain representations of $\ig$ may not be lifted to representations of
  $\iG$ and drops them from the set of boundary labels.
  
% ------------------------------------------------------------------------
% ------------------------------------------------------------------------
% ------------------------------------------------------------------------

\section{Proof of bundle triviality}
In this appendix we will prove that all complexified vector bundles
$G\times_{\iG} V_{R}^{\Complex } $ over the base manifold $G/\iG $ associated
to representations $V_{R}$ of $\iG $ are trivial. Here, $G$ is any
simple simply-connected compact Lie group and $\iG $ the subgroup invariant
under a diagram automorphism. All possible cases are summarized in
table~\ref{tb:Overview} on page~\pageref{tb:Overview}.
\smallskip 

Before we start with the actual proof, let us note that 
representations $V_{R}$ which arise by restricting  
representations of $G$ to $\iG $ always lead to trivial bundles. We
will use this extensively to proof the triviality of all other
bundles.
\smallskip

The proof can be devided into five parts. We will first present these 
five main statements and then enter the detailed discussion of the 
single steps.
\bigskip 

\noindent
{\bf Statement 1:} Consider the K-ring $K_{\Complex } (G/\iG )$ of complex 
vector bundles over the base manifold $G/\iG $. The map $\mc{K}:V_{R}\to K 
(G\times_{\iG} V_{R}^{\Complex })$ which sends a representation $V_{R}$ of 
$\iG $ to the K-class of its associated complexified vector bundle, is a 
\textit{ring homomorphism} from the representation ring ${\rm Rep} (\iG )$ 
to $K_{\Complex } (G/\iG )$.
\bigskip

\noindent
{\bf Statement 2:} The representation ring of $\iG $ is a polynomial ring 
on the fundamental representations, ${\rm Rep} (\iG)=\Integer [V_{\omega 
_{1}},\dots ,V_{\omega _{r}}]$, $r=\rank \iG $.\\
Therefore any K-class of a bundle $G\times_{\iG} V_{R}^{\Complex } $ can be
expressed as a polynomial in the K-classes of $G\times_{\iG } 
V_{\omega_{i}}^{\Complex } $.
\bigskip 

\noindent
{\bf Statement 3:} All complexified vector bundles associated to
fundamental representations of $\iG $ have trivial K-class, $K
(G\times_{\iG} V_{\omega _{i}}^{\Complex })=0$, i.e.\ all these 
bundles are stably equivalent to trivial bundles.\\
From the previous remark it follows then that all bundles $G
\times_{\iG} V_{R}^{\Complex } $ are stably equivalent to trivial 
bundles.
\bigskip

\noindent
{\bf Statement 4:} Two stably equivalent 
complex vector bundles of rank $d$ over a
base manifold of dimension $n$ satisfying 
$2d\geq n$ are isomorphic.
\bigskip

\noindent
{\bf Statement 5:}
All representations $V_{R}$ that are not a restriction of
a representation of $G$ obey the inequality
\begin{equation}\label{eq:rankdimension}
2\cdot \dim V_{R} \geq \dim G/\iG \ \ .
\end{equation}
We had seen that all bundles are stably equivalent to trivial bundles,
from the last two statements we can thus conclude that all bundles are 
trivial. This ends the main line of argumentation. Note that it was 
important that we considered complexified vector bundles, otherwise 
there would appear a much stronger inequality in statement 4 which in 
many cases could not be fulfilled any more.
\medskip 

Let us now take a closer look at the single statements. The first
statement follows from the fact that the bundles associated to the
tensor product of two representations $V_{R}\otimes V_{R'}$ is the
tensor product of the associated bundles,
\[
G\times_{\iG} (V_{R}\otimes V_{R'}) \ \simeq \ ( G\times_{\iG} V_{R}
)\ \otimes_{G/\iG }\ ( G\times_{\iG} V_{R'}) \ \ .
\]
Statement 2 is a structure theorem which can be found e.g. in
\cite[Theorem~23.24]{FultonHarris}.
\smallskip

The third statement is much more technical. We have to check its
validity case by case. We know that restrictions of
representations of $G$ give rise to trivial bundles and thus to
trivial K-classes. Studying the appearance of fundamental
representations in the decomposition of $G$-representations, we deduce
in an inductive way that all fundamental representations of $\iG $
correspond to bundles of trivial K-classes.
Let us discuss the way it works in an example.
\smallskip

$B_{3}\embin D_{4}$: The fundamental representations of $B_{3}$ have
the dimensions 7, 8 and 21. The first fundamental representation of
$D_{4}$ is 8-dimensional and decomposes as $8\to 7+1$. The
corresponding bundle is trivial, as well as the bundle associated to
the trivial representation 1, hence the bundle associated to the
7-dimensional representation is stably equivalent to a trivial bundle.
The next fundamental representation of $D_{4}$ decomposes as $28\to
21+7$ so that we find by an analogous argument as above that the
bundle associated to the fundamental 21 of $B_{3}$ is stably equivalent
to a trivial bundle. The remaining 8-dimensional fundamental representation of
$B_{3}$ is the restriction of one of the 8-dimensional representations
of $D_{4}$ and thus gives rise to a trivial bundle.\smallskip

In a similar way we will prove statement 3 for all cases at the end of
this section. But before
we do so, we want to discuss the last two statements. 
After we have shown that the fundamental and thus all representations
give rise to bundles which are stably equivalent to trivial bundles,
we want to show that they are actually really trivial.
Statement 4 is a theorem that can be found e.g.\ in \cite[Theorem
9.1.5]{Husemoller} which tells us that ``s-equivalent'' and ``isomorphic''
have the same meaning if the rank of the bundles is high enough.
The last statement 5 ensures that all our bundles indeed comply with 
this requirement, hence they are trivial. To prove this statement
we determine for all cases the lowest-dimensional representation
not arising as a restriction of a $G$-representation. We show then that
these satisfy the inequality \eqref{eq:rankdimension}. 

As an example take again $B_{3}\embin D_{4}$. The lowest dimensional
non-trivial representation has dimension 7 which is not a restriction
of a representation of $D_{4}$. The base manifold has dimension $\dim
D_{4}-\dim B_{3}=28-21=7$. As $2\cdot 7=14\geq 7$, the inequality holds.

We will now show the validity of this statement for all cases together
with statement 3 in a case-by-case study. \bigskip 

\begin{itemize}
\item $B_{n}\embin A_{2n}\quad [SO (2n+1)\subset SU (2n+1)]$:\\
Decomposition of fundamental representations of $A_{2n}$ (Dynkin label
notation):
\[
\setlength{\arraycolsep}{0mm}
\begin{array}{p{8mm}p{5mm}p{18mm}cp{8mm}p{9mm}c}
$(0\dotfill$&\multicolumn{2}{c}{$1\dotfill
0)$}&\ \ \ \longrightarrow\ \ \ &$(0\dotfill$&$1\dotfill 0)$&\quad\quad  i<n\\
&$i$& & & &$i$ \\[1mm]
\multicolumn{2}{l}{$(0\dotfill$} &$1\dotfill 0)$&\ \ \longrightarrow\
\ &\multicolumn{2}{l}{$(0\dotfill 2)$}\\
& &$n$& & & 
\end{array}
\]
All fundamental weights of $SO (2n+1)$ are restrictions of fundamental
representations of $SU (2n+1)$. $\Rightarrow $~{\bf Statement~3}

The lowest dimensional representation not obtained from a restriction
is the representation $(20\dots 0)$ of $B_{n}$ with dimension
$2n^{2}+3n$.
The base manifold has dimension $\dim SU (2n+1) - \dim SO
(2n+1)=2n^{2}+3n$, so inequality \eqref{eq:rankdimension}
holds. $\Rightarrow $~{\bf Statement~5}

\item $C_{n}\embin A_{2n-1}\quad [Sp (2n)\subset SU (2n)]$:\\
Decomposition of fundamental weights of $A_{2n-1}$:
\[
\setlength{\arraycolsep}{0mm}
\begin{array}{p{13mm}p{20mm}cp{13mm}p{10mm}cp{11mm}p{12mm}c}
\multicolumn{2}{l}{$(1\dotfill 0)$}&\ \ \ \longrightarrow\ \ \ 
&\multicolumn{2}{l}{$(1\dotfill 0)$}\\[1mm]
\multicolumn{2}{l}{$(01\dotfill 0)$}&\longrightarrow 
&\multicolumn{2}{l}{$(01\dotfill 0)$}&\ \oplus\ 
&\multicolumn{2}{l}{$(0\dotfill 0)$}\\[1mm]
\multicolumn{2}{l}{$(001\dotfill 0)$}&\longrightarrow 
&\multicolumn{2}{l}{$(001\dotfill 0)$}&\oplus 
&\multicolumn{2}{l}{$(1\dotfill 0)$}\\[3mm]
$(0\dotfill $&$1\dotfill 0)$&\longrightarrow 
&$(0\dotfill$&$1\dotfill 0)$&\oplus 
&$(0\dotfill$&$1\dotfill 0)$&\ \oplus \dots \\
 &$i$& & &$i$& & &\makebox[2mm]{$i-2$}& \\
\multicolumn{2}{c}{(i\leq n)} & 
&\multicolumn{6}{c}{\dots \oplus 
\left\{\begin{array}{ll}\makebox[23mm]{$(1\dotfill 0)$}&
\ \ \ i\text{ odd}\\
\makebox[23mm]{$(0\dotfill 0)$}&\ \ \ i\text{ even}
\end{array} \right. 
}
\end{array}
\]
Proceeding inductively, we see that all bundles associated to
fundamental representations of $C_{n}$ have trivial
K-class. $\Rightarrow $~{\bf Statement~3}

The lowest dimensional representation that cannot be 
obtained from a restriction
is $(010 \dots 0)$ and has dimension $2n^{2}-n-1$. The base manifold
has dimension $\dim SU (2n)-\dim Sp (2n)=2n^{2}-n-1$. 
$\Rightarrow $~{\bf Statement~5}

\item $B_{n-1}\embin D_{n}\quad [Spin (2n-1)\subset Spin (2n)]$:\\
Decomposition of fundamental representations of $D_{n}$:
\[
\setlength{\arraycolsep}{0mm}
\begin{array}{p{13mm}p{10mm}cp{13mm}p{10mm}cp{11mm}p{12mm}}
\multicolumn{2}{l}{(10\dotfill0)}&\ \ \ \longrightarrow\ \ \ 
&\multicolumn{2}{l}{(10\dotfill0)}&\ \ \oplus \ \ 
&\multicolumn{2}{l}{(0\dotfill0)}\\[1mm]
\multicolumn{2}{l}{(01\dotfill0)}&\ \ \ \longrightarrow\ \ \ 
&\multicolumn{2}{l}{(01\dotfill0)}&\ \ \oplus \ \ 
&\multicolumn{2}{l}{(10\dotfill0)}\\[3mm]
$(0\dotfill$&$1\dotfill0)$&\longrightarrow
&$(0\dotfill$&$1\dotfill0)$& \oplus 
&$(0\dotfill$&$1\dotfill0)$\\
&\makebox[0cm][l]{$i\ \ (i<n-1)$}& & &$i$& & &\makebox[2mm]{$i-1$}\\[2mm]
\multicolumn{2}{l}{\begin{array}{p{23mm}}
$(0\dotfill10)$\\[1mm] 
$(0\dotfill01)$
\end{array}}&
\begin{array}{c}
\rotatebox{-30}{$\longrightarrow$} \\[1mm] \rotatebox{30}{$\longrightarrow$}
\end{array}&
\multicolumn{2}{l}{(0\dotfill 1)}
\end{array}
\]
$\Rightarrow $~{\bf Statement~3}

The lowest dimensional non-trivial representation of $B_{n-1}$ is
$(10\dots 0)$ and has dimension $2n-1$. The dimension of the base
manifold is $\dim Spin (2n)-\dim Spin
(2n-1)=2n-1$. $\Rightarrow $~{\bf Statement~5}

\item $F_{4}\embin E_{6}$:\\
Decomposition of representations of $E_{6}$:
\begin{eqnarray*}
(000010)&\longrightarrow & (0001)\oplus (0000)\\
(000001)&\longrightarrow & (1000)\oplus (0001)\\
(000100)& \longrightarrow & (0010)\oplus (1000)\oplus (0001) 
\end{eqnarray*}
So we see that the three fundamental representations $(1000)$,
$(0010)$, $(0001)$ of $F_{4}$ give rise to bundles of vanishing
K-class. From the tensor product
\[
(0001)\otimes (1000)\ =\ (0001)\oplus (0010)\oplus (1001)
\]
we can deduce that the same is valid for $(1001)$.

Now we look at the decomposition
\[
(001000)\ \longrightarrow\ (0100)\oplus (1001)\oplus (0010)\oplus
(0010)\oplus (1000)
\]
and we find a vanishing K-class also for the fourth 
fundamental representation $(0100)$. $\Rightarrow $~{\bf Statement~3}

The lowest dimensional representation of $F_{4}$ is $(0001)$ and has
dimension 26. The dimension of the base manifold is $\dim E_{6}-\dim
F_{4}=26$. $\Rightarrow $~{\bf Statement~5} 

\item $G_{2}\embin D_{4}$:\\
Decompositions of representations of $D_{4}$:
\begin{eqnarray*}
(1000)& \longrightarrow & (01)\oplus (00)\\
(0100)& \longrightarrow & (10)\oplus (01)\oplus (01)
\end{eqnarray*} 
$\Rightarrow $~{\bf Statement~3}

The lowest dimensional non-trivial representation of $G_{2}$ is
$(01)$ and has dimension 7. The base manifold has dimension $\dim
D_{4}-\dim G_{2}=28-14=14$. $\Rightarrow $~{\bf Statement~5}
\end{itemize}
\end{appendix}

% Take care of this newpage!!!
\newpage

\providecommand{\href}[2]{#2}\begingroup\raggedright\endgroup

%\bibliographystyle{$HOME/texstyles/JHEP-2}
%\bibliography{$HOME/bibliography/bibliography}

\end{document}